\documentclass{aa}       
\usepackage{graphicx}
\usepackage{amssymb}
\usepackage{amsmath}
\usepackage{txfonts}

\begin{document}

\title{A supermassive binary black hole in the quasar 3C\,345}

\author{
A.P. Lobanov\inst{1}
\and J. Roland\inst{2}
}

\authorrunning{Lobanov \& Roland}
\titlerunning{Binary black hole in 3C\,345}

\institute{Max-Planck-Institut f\"ur Radioastronomie, 
           Auf dem H\"ugel 69, Bonn 53121, Germany  
           \and 
           Institut d'Astrophysique, bd. Arago 98\,bis,
           75014 Paris, France}

\offprints{A.P. Lobanov, \email{alobanov@mpifr-bonn.mpg.de}}

\date{Received <date> / Accepted <date>}

\abstract{Radio loud active galactic nuclei present a remarkable
  variety of signs indicating the presence of periodical processes
  possibly originating in binary systems of supermassive black holes,
  in which orbital motion and precession are ultimately responsible
  for the observed broad-band emission variations, as well as for the
  morphological and kinematic properties of the radio emission on
  parsec scales. This scenario, applied to the quasar
  3C\,345, explains the observed variations of radio and optical
  emission from the quasar, and reproduces the structural variations
  observed in the parsec-scale jet of this object.  The binary system
  in 3C\,345 is described by two equal-mass black holes with masses of
  $\approx7.1\times 10^8{\mathrm M}_{\sun}$ separated by $\approx
  0.33$\,pc and orbiting with a period $\sim 480$\,yr. The orbital
  motion induces a precession of the accretion disk around the primary
  black hole, with a period of $\approx 2570$\,yr.  The jet plasma is
  described by a magnetized, relativistic electron-positron beam
  propagating inside a wider and slower electron-proton jet. The
  combination of Alfv\'en wave perturbations of the beam, the orbital
  motion of the binary system and the precession of the accretion disk
  reproduces the variability of the optical flux and evolution of the
  radio structure in 3C\,345. The timescale of quasi-periodic flaring
  activity in 3C\,345 is consistent with typical disk instability
  timescales. The present model cannot rule out a small-mass orbiter
  crossing the accretion disk and causing quasi-periodic flares.
}

\maketitle

\section{Introduction}
\label{sc:sect1}

Most of the radio loud active galactic nuclei (AGN) exhibit emission
and structural variability over the entire 
electromagnetic spectrum, on timescales ranging from several hours to
several years and on linear scales of from several astronomical units to
several hundreds of parsecs.  In AGN with prominent relativistic jets,
this variability appears to be related to the observed morphology and
kinematics of the jet plasma on parsec scales (e.g., Zensus et al.
\cite{zen02}; Jorstad et al. \cite{jor+01}).  In a number of radio-loud
AGN, enhanced emission regions ({\it components}) embedded in the jet
move at superluminal speeds along helical trajectories (Zensus,
Krichbaum \& Lobanov \cite{zkl95}, \cite{zkl96}; Zensus \cite{zen97}), and the
ejection of such components is often associated with optical and/or
$\gamma$-ray outbursts.

In a growing number of cases, explanation of the observed nuclear
variability and structural changes in parsec-scale jets is linked to
the presence of supermassive binary black hole (BBH) systems in the
centers of radio loud AGN.  The BBH model was originally formulated by
Begelman, Blandford \& Rees (\cite{bbr80}), and it was studied
subsequently in a number of works (e.g., Polnarev \& Rees \cite{pr94};
Makino \& Ebisuzaki \cite{me96}; Makino \cite{mak97}; Ivanov,
Papaloizou \& Polnarev \cite{ipp99}).  The BBH model postulates that
an accretion disk (AD) exists around at least one of the two black
holes (typically, around the more massive one), and the resulting
variability and structural changes are determined by the dynamic
properties of the disk itself and the BBH--AD system (this
includes the disk and black hole precession, orbital motion, passages
of the secondary component through the AD).
  
The most celebrated examples of successful application of the BBH
model include \object{OJ\,287} (Sillanp\"a\"a, Haarala \& Valtonen
\cite{shv88}; Valtaoja et al. \cite{val+00}), \object{3C\,273} and
\object{M\,87} (Kaastra \& Roos \cite{kr92}), \object{1928$+$738}
(Roos, Kaastra \& Hummel \cite{rkh93}), \object{Mrk\,501} (Rieger \&
Mannheim \cite{rm00}) and \object{PKS\,0420$-$014} (Britzen et al.
\cite{bri+01}).  The epochs and structure of major outbursts in
\object{OJ\,287} have been suggested to result from passages of the
secondary black hole of a binary pair in \object{OJ\,287} through the
accretion disk around the primary during the orbit (Lehto \& Valtonen
\cite{lv96}).  The component ejection epochs are correlated with the
optical outbursts, and the component trajectories show evidence for a
helical morphology of the jet in \object{OJ\,287} (Vicente, Charlot \&
Sol \cite{vcs96}).  Several jet components observed in
\object{3C\,345} also move along distinct helical paths (Steffen et
al.  \cite{ste+95}; Lobanov 1996, hereafter \cite{L96}).  Lobanov \&
Zensus (1999, hereafter \cite{LZ99}) have shown that shocks are not
likely to play a significant role in the dynamics and emission of
parsec-scale regions in \object{3C\,345}.  Recently, the presence of
binary black hole systems (albeit with extremely short precession
periods and small orbital separations) has been suggested for 3C\,279
(Abraham \& Carrara \cite{ac98}), \object{3C\,273} (Abraham \& Romero
\cite{ar99}) and \object{3C\,345} (Caproni \& Abraham \cite{ca04}),
from the observed periodic variations of the component ejection angle.
  
The observed helical trajectories of jet components are a strong
indication of precession or perturbation of the jet flow.  The
dynamics and emission of such perturbed outflows have been explained
in the framework of the two-fluid model (Sol, Pelletier \& Asseo
\cite{spa89}, Hanasz \& Sol \cite{hs96}) describing the structure and
emission of the jet in terms of an ultra-relativistic
electron-positron ($e^{\pm}$) beam with $\Gamma_\mathrm{ b} \approx
10$ surrounded by a slower, electron-proton ($e^{-}p$) jet moving at a
speed $\beta_\mathrm{ j} \approx 0.4\,c$.
  
Observational evidence for the two-fluid model is reviewed in Roland
\& Hetem (\cite{rh96}). The presence of two fluids in extragalactic
outflows was first inferred in the large-scale jet of Cygnus~A
(Roland, Pelletier \& Muxlow \cite{rpm88}). The two-fluid model is
supported by $\gamma$-ray observations of MeV sources (Roland \&
Hermsen \cite{rh95}; Skibo, Dermer \& Schlickeiser \cite{sds97}) and
polarized structures in the compact radio jet of \object{1055$+$018}
(Attridge, Roberts \& Wardle \cite{arw99}).  The X-ray and
$\gamma$-ray observed in \object{Cen~A} may be formed by the $e^{\pm}$
beams (Marcowith, Henry \& Pelletier \cite{mhp95}; Marcowith, Henri \&
Renaud \cite{mhr98}). Direct detections of fast and slow jet speeds
have been reported for \object{Cen~A} (Tingay et al.  \cite{tin+98})
and \object{M\,87} (Biretta, Sparks \& Macchetto \cite{bsm99}).
Subluminal speeds, possibly related to the $e^{-}p$ plasma, have been
observed in the double-sided jets in \object{3C\,338} (Feretti et al.
\cite{fer+93}). The two-fluid scenario has been used to explain the
morphology and velocity field in the large-scale jet in
\object{3C\,31} (Laing \& Bridle \cite{lb02}, \cite{lb04}).
  
Orbital motion and disk precession in the BBH--AD system perturb (but
not disrupt) the $e^{\pm}$ beam, which leads to the complex
variability and kinematic patterns observed in the compact jets.  On
scales of up to several hundreds of parsecs, the $e^{\pm}$ beam is not
disrupted by Langmuir turbulence (Sol, Pelletier \& Asseo
\cite{spa89}) and Alfv\'en and whistler waves (Pelletier \& Sol
\cite{ps92}; Achatz \& Schlickeiser \cite{as93}).  No strong
Kelvin-Helmholtz instability should form on these scales, despite the
significant transverse stratification of the outflow (Hanasz \& Sol
\cite{hs96}).  Kinematic and emission properties of perturbed beams
have been calculated (Roland, Teyssier \& Roos \cite{rtr94}; Desringe
\& Fraix-Burnet \cite{df97}), and it has been shown that the observed
variability and non-linear trajectories of jet components may indeed
be caused by a perturbed $e^{\pm}$ plasma.  In a more general
approach, solutions for the restricted three-body problem (Laskar
\cite{las90}; Laskar \& Robutel \cite{lr95}) have been applied for
deriving the parameters of the perturbed beam and precessing accretion
disk in the BBH system in the quasar \object{PKS\,0420$-$014} (Britzen
et al. \cite{bri+01}).
  
In this paper, the dynamics of the BBH--AD system and the two-fluid
model are combined together in order to construct a single framework
for explaining optical flaring activity, kinematic properties and
internal structure of the compact parsec-scale jets.  The combined
model is developed to describe the quasar \object{3C\,345} . It is
presented in Sect.~\ref{sc:sect2}.  Section~\ref{sc:sect3} provides
an overview of the basic properties of the optical and radio emission from
\object{3C\,345}. In sections~\ref{sc:sect4}--\ref{sc:sect5}, the
model is applied to describe the evolution of the optical emission and
radio jet in \object{3C\,345} after a strong flare in 1992 that
resulted in the appearance of a new superluminal jet component C7.

Throughout the paper, the Hubble constant $H_{0}=70\,h\,$km\,
s$^{-1}$\,Mpc$^{-1}$, deceleration parameter $q_{0}=0.5$, and negative
definition of the spectral index, $S\propto\nu^{-\alpha}$ are used. For
\object{3C\,345} ($z=0.595$, Hewitt \& Burbidge \cite{hb93}), the adopted
cosmological parameters correspond to a luminosity distance
$D_\mathrm{ L}=1.99h^{-1}$\,Gpc. The corresponding linear scale is
$3.79h^{-1}$\,pc\,mas$^{-1}$. A proper motion of 1\,mas/year
translates into an apparent speed of $19.7h^{-1}c$. Subscripts ``b'',
``j'', ``c'', and ``m'' are introduced to identify quantities related
to the beam, jet, moving plasma condensation, and magnetic field,
respectively.

\section{The binary black hole model}
\label{sc:sect2}

\begin{figure}[t]
\centerline{
\includegraphics[width=0.45\textwidth, bb =0 0 815 568, clip=true]
{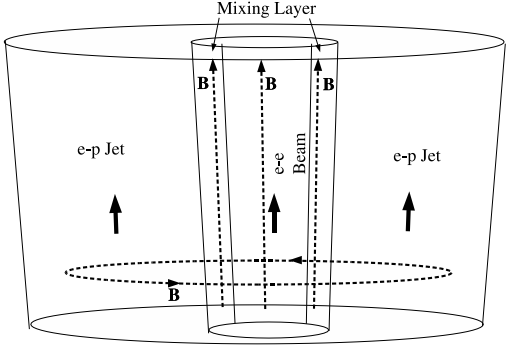}}
\caption{Two-fluid model for relativistic outflows. A fast,
  relativistic $e^{\pm}$ beam is surrounded by a slower, possibly
  thermal, $e^{-}p$ outflow, with a mixing layer forming between the
  beam and the jet.  The magnetic field, $B$, is parallel to the flow
  in the beam and in the mixing layer, and the field is toroidal in the
  jet. The magnetic lines are perturbed and the perturbation
  propagates outwards at the Alfv\'en speed, $V_{\rm A}$.}
\label{fg:twofluid}
\end{figure}

The binary black hole model provides a general formalism for
calculating the emission and kinematics of an outflow originating from
a BBH-AD system.  The two-fluid description of the outflow is
adopted (Fig.~\ref{fg:twofluid}), with the following assumptions:

\begin{itemize}

\item[1.]~The outflow consists of an $e^{-}p$ plasma (hereafter {\em
the jet}) moving at mildly relativistic speed $\beta_\mathrm{ j} \le
0.4\,c$ and an $e^{\pm}$ plasma (hereafter {\em the beam}) moving at
highly relativistic speed (with corresponding Lorentz factors
$\Gamma_\mathrm{ b}\sim 10$). Both beam and jet may be transversely
stratified.

\item[2.]~The magnetic field lines are parallel to the flow in the
  beam and the mixing layer, and the field is toroidal in the jet
  (Fig.~\ref{fg:twofluid}). The energy densities of the beam, jet
  and magnetic field are related as follows: $\varepsilon_\mathrm{
    j}\gg \varepsilon_\mathrm{ b} \approx \varepsilon_\mathrm{ m}$.
  The magnetic field lines are perturbed, and the perturbation
  propagates downstream at the Alfv\'en speed, $V_\mathrm{ A}$.

\end{itemize}

In the two-fluid model, the $e^{-}p$ jet carries most of the mass and
kinetic power released on kiloparsec scales (large scale jets,
extended lobes, hot spots).  The $e^{\pm}$ beam propagates in a
channel inside the jet, and it is mainly responsible for the formation
of superluminal jet components observed on parsec scales.  The beam
interior is assumed to be homogeneous, ensuring that the beam rotation
does not affect the observed evolution.  Assuming a pressure
equilibrium between the beam and the jet, a typical size of the beam
is $R_\mathrm{ b} \approxeq 0.1R_\mathrm{ j}$, which is of the order
of 0.1\,pc (Pelletier \& Roland \cite{pr89}).  For AGN at distances of
the order of 1\,Gpc, this translates into an angular dimension of
$\sim$30 microarcseconds, and the beam remains transversely unresolved
for very long baseline interferometry (VLBI) observations at
centimeter wavelengths.

\subsection{Geometry of the model}

\begin{figure}[t]
\centerline{
\includegraphics[width=0.48\textwidth, bb =0 0 414 407, clip=true]
{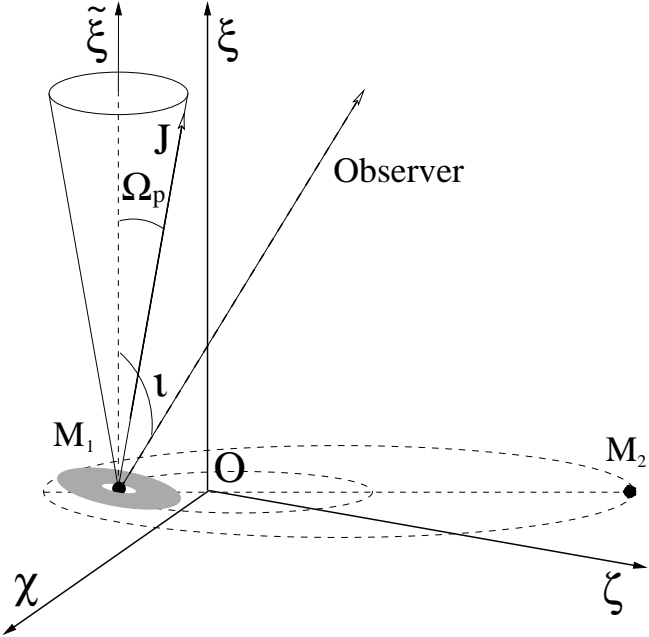}}
\caption{Geometry of the BBH model. M$_1$ and M$_2$ denote the locations of
  two black holes orbiting around each other in the plane ($\chi$O$\zeta$).
  The coordinate system is referenced to the center of mass of the
  binary system (O). The accretion disk around the black hole M$_1$
  inclined at an angle $\Omega_\mathrm{ p}$ with respect to the orbital plane,
  such that the $\Tilde{\xi}$ axis (parallel to the $\xi$ axis) is the
  precession axis.  The momentary direction of the beam axis is given by
  the vector $\vec{{\mathrm M}_1 {\mathrm J}}$.}
\label{fg:scheme}
\end{figure}

Basic geometry of the model is illustrated in Fig.~\ref{fg:scheme}.
Two black holes M$_1$ and M$_2$ orbit each other in the plane
($\chi$O$\zeta$). The center of the mass of the binary system is in O.  An
accretion disk around the black hole M$_1$ is inclined at an angle
$\Omega_\mathrm{ p}$ with respect to the orbital plane. The angle $\Omega_\mathrm{ p}$ is
the opening half-angle of the precession cone. The disk precesses
around the $\Tilde{\xi}$ axis, which is parallel to the $\xi$-axis. The
momentary direction of the beam axis is then given by the vector
$\vec{{\mathrm M}_1 {\mathrm J}}$. The line of sight makes an angle $\iota$
with the precession axis.

\subsection{Stability and dissipation of the beam}

The beam exists inside the jet for as long as its density
\begin{equation}
n_\mathrm{ b} \ge \varepsilon_\mathrm{ j} / (\Gamma_\mathrm{ b}^2 m_\mathrm{ e} c^2)\,.
\end{equation}
For a typical parsec-scale jet, this implies $n_\mathrm{ b} \ga 10 e^{-}\,\mathrm{cm}^{-3}$.  The beam moves along magnetic field lines in the
$e^{-}p$ jet if the magnetic field is larger than the critical value
\begin{equation}
B_\mathrm{ crit} = (4 \pi n_\mathrm{ mix} m_\mathrm{ e} c^2)^{1/2}\,
\end{equation} 
where $n_\mathrm{ mix}\approx n_\mathrm{ j}$ is the density of the
plasma in the beam--jet mixing layer (Pelletier, Sol \& Asseo
\cite{psa88}). The magnetic field in the jet rapidly becomes toroidal
(Pelletier \& Roland \cite{pr90}).  The beam is not disrupted by
Kelvin-Helmholtz perturbations if the magnetic field is larger than
\begin{equation}
B_\mathrm{ KH} = (2 \pi n_\mathrm{ b} \Gamma_\mathrm{ b}^2 m_\mathrm{ e}
c^2)^{1/2}
\end{equation} 
(Pelletier \& Roland \cite{pr89}). 
A typical density ratio in parsec-scale jets is
$\eta=(n_\mathrm{ b}/n_\mathrm{ j}) \approxeq 0.01$, which implies
$B_\mathrm{ crit} \sim B_\mathrm{ KH} \sim 50$\,mG.  Sol, Pelletier \&
Asseo (\cite{spa89}) have demonstrated that the beam does not suffer
  from inverse Compton and synchrotron losses on parsec scales.
  Relativistic bremsstrahlung, ionization and annihilation losses are
  also insignificant on these scales. Diffusion by turbulent Alfv\'en
  waves is not expected for beams with $\Gamma_\mathrm{ b} \sim 10$.

\subsection{Kinematics of the perturbed beam}

The relativistic plasma injected into the beam in the vicinity of the black
hole M$_1$ moves at a speed characterized by the bulk Lorentz factor
$\Gamma_\mathrm{ c}$. The condensation follows the magnetic field lines. The
perturbation of the magnetic field propagates at the Alfv\'en speed
$V_\mathrm{ A} = B (4 \pi m_\mathrm{ e} n_\mathrm{ b})^{-0.5}$.  
For highly energetic beams, the classical density $\rho_0 = m_\mathrm{e}
  n_\mathrm{b}$ is substituted with the enthalpy $\mu = \rho + P/c^2$, where $P$
  is the pressure, $\rho = \rho_0 (1 + \varepsilon/c^2)$ is the relativistic
  density and $\varepsilon$ is the specific internal energy.  Taking into
  account relativistic corrections, and recalling that the enthalpy becomes
  $\Gamma_\mathrm{b}(\rho + P/c^2)$ in a plasma moving at a bulk speed
  $\beta_\mathrm{ b} = (1 - 1/\Gamma_\mathrm{b}^2)^{0.5}$, the resulting phase
  speed of the Alfv\'en perturbation with respect to the beam plasma is
  $V_\mathrm{ A,rel} = \sqrt{1/(1+\Gamma_\mathrm{ b}/V_\mathrm{ A}^2)}$.  
Thus, for a typical $e^{\pm}$ beam propagating at
$\Gamma_\mathrm{ b}\ga 10$ inside a jet with $B\ga B_\mathrm{ crit}$, the
relativistic Alfv\'en speed is $\approx 0.1\,c$.

The observed trajectory of the beam is affected primarily by the
orbital motion in the BBH system and the precession of the AD around
the primary BH. 
The jet structure on parsec scales is
mainly determined by the precession. The orbital motion introduces a
small ($\la 0.1$\,mas) oscillation of the footpoint of the jet, and it
is not immediately detectable in VLBI images with a typical resolution
of $\approx 1$\,mas.  
Consider first the effect of the
precession of the accretion disk. The coordinates of the component
moving in the perturbed beam are given by
\begin{equation}
\left\{
\begin{split}
\Tilde{\chi}  &  =R(\xi)\cos(\omega_\mathrm{ p}t-k_\mathrm{ p}\xi + \phi_\mathrm{ p}) \\
\Tilde{\zeta}  &  =R(\xi)\sin(\omega_\mathrm{ p}t-k_\mathrm{ p}\xi + \phi_\mathrm{ p}) \\
\Tilde{\xi}  &  = \xi = \xi(t)\,,
\end{split}
\right.
\label{eq:xyz_prime}
\end{equation}
where $R(\xi)$ is the perturbation amplitude, $\phi_\mathrm{ p}$ is the
initial precession phase, $T_\mathrm{ p}$ is the precession period,
$\omega_\mathrm{ p}=2\pi/T_\mathrm{ p}$ and $k_\mathrm{ p}=\omega_\mathrm{ p}/V_\mathrm{
A}$. The time $t$ is measured in the rest frame of the beam.  The
tilde in $\Tilde{\chi}$ and $\Tilde{\zeta}$ indicates that the
orbital motion has not yet been taken into account (here the axes $\Tilde{\chi}$
and $\Tilde{\zeta}$ are assumed to be parallel to the axes $\chi$ and
$\zeta$, respectively; see Fig.~\ref{fg:scheme}).  Obviously,
$\Tilde{\xi} = \xi$, and the tilde can be omitted in all
expression with $\xi$.  The $R(\xi)$ can be chosen to be
\begin{equation}
R(\xi)=\frac{R_{0}\xi}{a+\xi}\,.
\label{eq:R_z}
\end{equation}
Here, $a=R_{0}/(2\tan\Omega_\mathrm{ p})$, and $R_0$ is the maximum amplitude
of the perturbation.

The form of the function $\xi(t)$ depends on the evolution of
the speed $\beta_\mathrm{ c}$ of the plasma condensation injected into
the perturbed beam. From Eq.~(\ref{eq:xyz_prime}), the components of
$\beta_\mathrm{ c}$ are:
\begin{equation}
\label{eq:beta_prime}
\left\{
\begin{split}
\beta_{\Tilde{\chi}}  = &\; \beta_{\xi} ({\mathrm d}R/{\mathrm d}\xi)
\cos(\omega_\mathrm{ p}t- k_\mathrm{ p}\xi + \phi_\mathrm{ p}) - \\
&  - R(\xi)\,(\omega - k \beta_\xi) \sin(\omega_\mathrm{ p}t-k_\mathrm{ p}\xi + \phi_\mathrm{ p}) \\
\beta_{\Tilde{\zeta}}  = &\; \beta_{\xi} ({\mathrm d}R/{\mathrm d}\xi)
\sin(\omega_\mathrm{ p}t -k_\mathrm{ p}\xi + \phi_\mathrm{ p}) + \\
& + R(\xi)\,(\omega - k \beta_\xi) \cos(\omega_\mathrm{ p}t-k_\mathrm{ p}\xi + \phi_\mathrm{ p})\\
\beta_{\Tilde{\xi}} = &\; \beta_\xi = {\mathrm d} \xi/{\mathrm d} t\,,
\end{split}
\right.
\end{equation}
so that
\begin{equation}
\label{eq:beta_c}
\beta^2_\mathrm{ c} = ({\mathrm d}\xi/{\mathrm d}t)^2 C_1 + R^2(\xi)\,(\omega_\mathrm{
p} - k_\mathrm{ p} {\mathrm d}\xi/{\mathrm d}t)^2 \,,
\end{equation}
where $C_1 = ({\mathrm d}R/{\mathrm d}\xi)^2 + 1$, and remembering that
$\beta_\mathrm{ c}^2 = \beta_{\Tilde{\chi}}^2 + \beta_{\Tilde{\zeta}}^2 + \beta_{\Tilde{\xi}}^2$.  Equation~(\ref{eq:beta_c}) is a quadratic equation with
respect to ${\mathrm d} \xi/{\mathrm d} t$, and it can be written in the form of
\begin{equation}
\label{eq:beta_quad}
A_2 ({\mathrm d} \xi/{\mathrm d} t)^2 + A_1 ({\mathrm d} \xi/{\mathrm d} t) + A_0 = 0\,,
\end{equation}
with $A_2 = 1+ R^2(\xi) k^2_\mathrm{ p} + ({\mathrm d} R/{\mathrm d} \xi)^2$, $A_1 = -2
R^2(\xi) \omega_\mathrm{ p} k_\mathrm{ p}$ and $ A_0 = R^2(\xi_\mathrm{ c})
\omega^2_\mathrm{ p} - \beta^2_\mathrm{ c}$.  The component speed $\beta_\mathrm{
c}$ can be conveniently expressed as a function of $\xi$ or $\xi(t)$
(it can also be viewed constant, in the simplest case).  The jet
side solution of Eq.~(\ref{eq:beta_c}) can be obtained for the
time $t(\xi)$ in the rest frame of the beam, under the condition of
initial acceleration of the beam (${\mathrm d} \xi/dt>0$, for $\xi\rightarrow
0_+$). The formal solution of Eq.~(\ref{eq:beta_quad}) is
\begin{equation}
\label{eq:beta_sol}
\begin{split}
\frac{{\mathrm d} \xi}{{\mathrm d} t}  =&\; \frac{1}{A_2} \{  R^2(\xi) \omega_\mathrm{ p}
k_\mathrm{ p} + \\
 &\; + \left[ R^2(\xi) k^2_\mathrm{ p} \beta^2_\mathrm{ c} +
(\beta^2_\mathrm{ c} - R^2(\xi) \omega^2_\mathrm{ p})(1 + 
\frac{{\mathrm d} R}{{\mathrm d} \xi})  \right]^{1/2}\}\,, \\
\end{split}
\end{equation}
and $t(\xi)$ is then given by
\begin{equation}
\label{eq:t_z}
\begin{split}
t(\xi) & =  t_0 + \int^{\xi}_{\xi_{0}} \frac{A_2}{k_\mathrm{ p}\omega_\mathrm{ p} 
R^2(\xi) + [A_2
\beta^2_\mathrm{ c} - \omega^2_\mathrm{ p} R^2(\xi) C_1]^{1/2}} {\mathrm d} \xi\,. \\
\end{split}
\end{equation}
In general, both
$\omega_\mathrm{ p}$ and $k_\mathrm{ p}$ can also vary. Their variations
should then be given with respect to $\xi$, in order to preserve
the solution given by Eq.~(\ref{eq:t_z}). Numerical methods should be used
otherwise.  The solution determined by Eq.~(\ref{eq:t_z}) can
be found for all $R(\xi)$ not exceeding the limit of
\begin{equation}
\label{eq:Rz_lim}
R(\xi) \le \beta_\mathrm{ c} \left[ \omega^2_\mathrm{ p}  - k^2_\mathrm{ p}
\beta^2_\mathrm{ c} \cos^2\psi(\xi)\right]^{-1/2}
\end{equation}
The beam pitch angle $\psi(\xi) = \arctan({\mathrm d}R/{\mathrm d}\xi)$ in
Eq.~(\ref{eq:Rz_lim}) is measured ``locally'' (as opposed to the
``global'' pitch angle $\psi_\mathrm{ g}(\xi) = \arctan[R(\xi_\mathrm{
    c})/\xi]$), and it
describes the evolution of the amplitude of the beam perturbation. For
a jet with constant beam pitch angle ($\psi(\xi) = \psi = const$), the
inequality \ref{eq:Rz_lim} can be rewritten as:
\begin{equation}
\label{z_lim}
\xi \le \beta_\mathrm{ c} \left[ \omega^2_\mathrm{ p} \tan^2\psi - k^2_\mathrm{ p} \beta^2_\mathrm{ c} \sin^2\psi
\right]^{-1/2}
\end{equation}
The speed $\beta_\mathrm{ c}$ and the local pitch angle $\psi$
should satisfy the condition:
\begin{equation}
\label{eq:Va_lim}
\beta_\mathrm{ c} \cos\psi \le \frac{V_\mathrm{ A}}{({V^2_\mathrm{ A} +1})^{0.5}}\,,
\end{equation}
If inequalities \ref{eq:Rz_lim}--\ref{eq:Va_lim} are satisfied, the
solution provided by Eq.~(\ref{eq:t_z}) can be used to represent
the form of the function $\xi(t)$ and describing the kinematics of the
beam.

\subsection{The binary system}

The trajectory of a superluminal feature travelling in the perturbed
beam is modified also by the orbital motion of the two black holes in
the binary system. This motion can be taken into account by solving
the restricted three-body problem for the BBH--AD system (Laskar
\cite{las90}, Laskar \& Robutel \cite{lr95}). The orbital plane is
$(\chi O \zeta)$, and the coordinate system is centered at the mass
center of the system. The orbit is given by $
r=p/[1+e\,\cos\varphi(t)]$, where $\varphi$ is the true anomaly, $p$
is the semi-latus rectum and $e$ is the numerical eccentricity of the
orbit.  The resulting coordinates ($\chi_1$, $\zeta_1$) of the black
hole M$_1$ are given by
\begin{equation}
\label{eq:M1_coord}
\left\{
\begin{split}
\chi_{1}(t)  &  =\frac{M_{2}}{M_{1}+M_{2}}\frac{p}{1+e\cos(\varphi(t))}
\cos(\varphi(t)) \\
\zeta_{1}(t)  &  =\frac{M_{2}}{M_{1}+M_{2}}\frac{p}{1+e\cos(\varphi(t))}
\sin(\varphi(t))
\end{split}
\right.
\end{equation}
The beam trajectory can be reconstructed from Eqs.~(\ref{eq:xyz_prime}) and (\ref{eq:M1_coord}), which gives
\begin{equation}
\label{eq:xyz}
\left\{
\begin{split}
\chi  &  =\Tilde{\chi} + \chi_{1}(t)\cos(\omega_\mathrm{ o}t-k_\mathrm{ o}\xi(t)
+ \phi_\mathrm{ o}) \\
\zeta  &  =\Tilde{\zeta} + \zeta_{1}(t)\sin(\omega_\mathrm{ o}t-k_\mathrm{ o}\xi(t)
+ \phi_\mathrm{ o}) \\
\xi  &  =\xi(t) 
\end{split}
\right.
\end{equation}
where $\phi_\mathrm{ o}$ is the initial orbital phase, $\omega_\mathrm{
o}=2\pi/T_\mathrm{ o}$, $k_\mathrm{ o} = \omega_\mathrm{ o}/ V_\mathrm{ A}$, and
$T_\mathrm{ o}$ is the orbital period.  This system of equations describes
completely the kinematics of the perturbed beam. Similarly to the
solution obtained for Eq.~(\ref{eq:beta_quad}), $\xi(t)$
can now be determined by forming $\beta_\mathrm{ c}^2$ from the coordinate
offsets given by Eq.~(\ref{eq:xyz}) and solving for ${\mathrm d} \xi/{\mathrm d} t$,
so that (for the jet side solution)
\begin{equation}
t(\xi) = t_0 + \int_{\xi_0}^{\xi} \frac{2 A_2}{({A_1}^2 - 4 A_1 A_0)^{1/2} -
A_1} {\mathrm d} \xi
\label{eq:tz_final}
\end{equation}
The coefficients $A_0$, $A_1$, and $A_2$ are given explicitly in
Appendix~II in Britzen et al. (\cite{bri+01}).

\subsection{Properties of the $t(\xi)$ solution}

The strongest effect on the form of the beam trajectory is produced by
three model parameters: the beam speed $\beta_\mathrm{ c}$, the Alfv\'en
speed $V_\mathrm{ A}$, and the precession period $T_\mathrm{ p}$. The orbital
period in the BBH system affects only the position of the footpoint of
the jet. Since VLBI measurements are made with respect to the
footpoint position, $T_\mathrm{ o}$ would not have an effect on fitting
the observed trajectories of the jet components (the orbital period is
constrained by the variability of the optical emission).  It is
easy to see that the remaining parameters ($R_0$, $\Omega_\mathrm{ p}$, $\iota$
and $\theta$) describe only the orientation and shape of the overall
envelope within which the perturbed beam is evolving.
The $t(\xi)$ solutions described by Eqs.~(\ref{eq:t_z}) and
\ref{eq:tz_final} can degenerate or become non-unique with respect
to $\beta_\mathrm{ c}$, $V_\mathrm{ A}$, $T_\mathrm{ p}$, if these parameters are
correlated. The combinations of $\beta_\mathrm{ c}$, $T_\mathrm{ p}$ 
and $V_\mathrm{ A}$ describing the same beam trajectory must not depend on the
rest frame time $t$ or beam location $\xi$. However, the solution given by
equation
\ref{eq:t_z}
implies that the beam trajectory would remain the same if
\begin{equation}
\label{eq:Tp_Va}
T_\mathrm{ p} = \frac{2 \pi R_0 \xi (\xi - V_\mathrm{ A} t)}
{(a+\xi)(\beta^2_\mathrm{ c} t^2 - \xi^2)^{1/2} V_\mathrm{ A}}\,.
\end{equation}
Obviously, Eq.~(\ref{eq:Tp_Va}) depends on both $t$ and $\xi$. The
dependence vanishes only when $t\rightarrow \xi/c$. The precession
period then becomes a complex number, and the resulting combination of
the parameters does not describe a physically meaningful situation.
This implies that the model does not become degenerate or non-unique with
respect to $\beta_\mathrm{ c}$, $T_\mathrm{ p}$ and $V_\mathrm{ A}$, and each physically 
possible combination of these parameters describes
a unique beam trajectory.

\subsection{Specific properties of the model}

The model introduces two new properties in the description of the
BBH--AD system: 1)~the Lorentz factor of the relativistic beam is time
variable; 2)~the initial perturbation of the magnetic lines in the
beam dissipates on a timescale $t_\mathrm{ beam}$. These modifications
are introduced in order to accommodate for the kinematic properties of
the jet in \object{3C\,345} in which the curvature of component trajectories
decreases at larger distances, and accelerated motions are strongly implied
by the observed trajectories and apparent speeds of several jet
components (\cite{LZ99}). The beam accelerates from $\Gamma_\mathrm{
  min}$ to $\Gamma_\mathrm{ max}$ over a rest frame time interval
$t_\mathrm{ acc}$, and
\begin{equation}
\Gamma_\mathrm{ c}(t) = \frac{\Gamma_\mathrm{ min} + \Gamma_\mathrm{ max} (t/t_\mathrm{ acc})}
{1 + t/t_\mathrm{ acc}}\;.
\label{eq:Gamma_t}
\end{equation}

The perturbation dissipates at a characteristic time $t_\mathrm{
  beam}$. The amplitude of the perturbation given by equation
  ({\ref{eq:R_z}) is modified by an attenuation term (characteristic
  damping time of the perturbation), so that
\begin{equation}
R(\xi)=\frac{R_{0}\xi}{a+\xi} \, e^{-t/t_\mathrm{ beam}}\;.
\label{eq:R_z_exp}
\end{equation}

The model takes account of an arbitrary source orientation, by
introducing a position angle, $\Theta_{\xi}$ of the $\xi$-axis in the
picture plane of the observer.

\subsection{Observed quantities}

Corrections for the relativistic motion of the beam must be made, in
order to enable calculations of physical quantities in the observer's
frame. This requires a knowledge of the angle $\theta(t)$ between the
instantaneous velocity vector of the component and the line of sight
(LOS).  Since the observed angular size of the BBH orbit is small
($\ll 1$\,milliarcsecond), the LOS can be assumed to lie in the
plane parallel to $(\chi O\xi)$, making an angle $\iota$ with the $\xi$
axis.  In this case, 
\begin{equation}
\label{eq:costheta}
\cos\theta(t)=\left( \frac{{\mathrm d}\chi}{{\mathrm d}t}\sin
      \iota+\frac{{\mathrm d}\xi}{{\mathrm d}t}\cos \iota\right)
      \frac{1}{\beta_\mathrm{ c}}
\end{equation}
in the observer's frame (Camenzind \& Krockenberger \cite{ck92}), with the resulting Doppler
beaming factor $\delta_\mathrm{ c}(t)=\{\Gamma_\mathrm{ c}[1-\beta_\mathrm{
c}\cos\theta(t)]\}^{-1}$ The observed flux density of the moving
component is
\begin{equation}
\label{eq:S_obs}
S_\mathrm{ c}=\frac{1}{D^2_\mathrm{ L}}\delta_\mathrm{ c}(t)^{1+\alpha_\mathrm{
    s}}(1+z)^{1+\alpha_\mathrm{ s}}\int_\mathrm{ c}
j_\mathrm{ c}{\mathrm d} V\,,
\end{equation}
where $z$ and $D_\mathrm{ L}$ are the redshift and luminosity distance of
the source, $j_\mathrm{ c}$ is the emissivity of the component and
$\alpha_\mathrm{ s}$ is the synchrotron spectral index. 
The shortening of time in the observer's frame is given by
\begin{equation}
\label{eq:t_obs}
t_\mathrm{ obs}=(1+z)\int_{0}^{t}\frac{{\mathrm d} t}{\Gamma_\mathrm{ c}(t)
\delta_\mathrm{ c}(t)}\,,
\end{equation}
and the component travels over the distance
\begin{equation}
\label{eq:dist_trav}
R_\mathrm{ trav} (t) =  (1+z)^{-1} \int_{0}^{t} \frac{\beta_\mathrm{ c}}
{1 - \beta_\mathrm{ c} \cos\theta(t)} {\mathrm d} t\,.
\end{equation}

\begin{figure*}
\centerline{
\includegraphics[width=0.95\textwidth, bb =6 9 773 593, clip=true]
{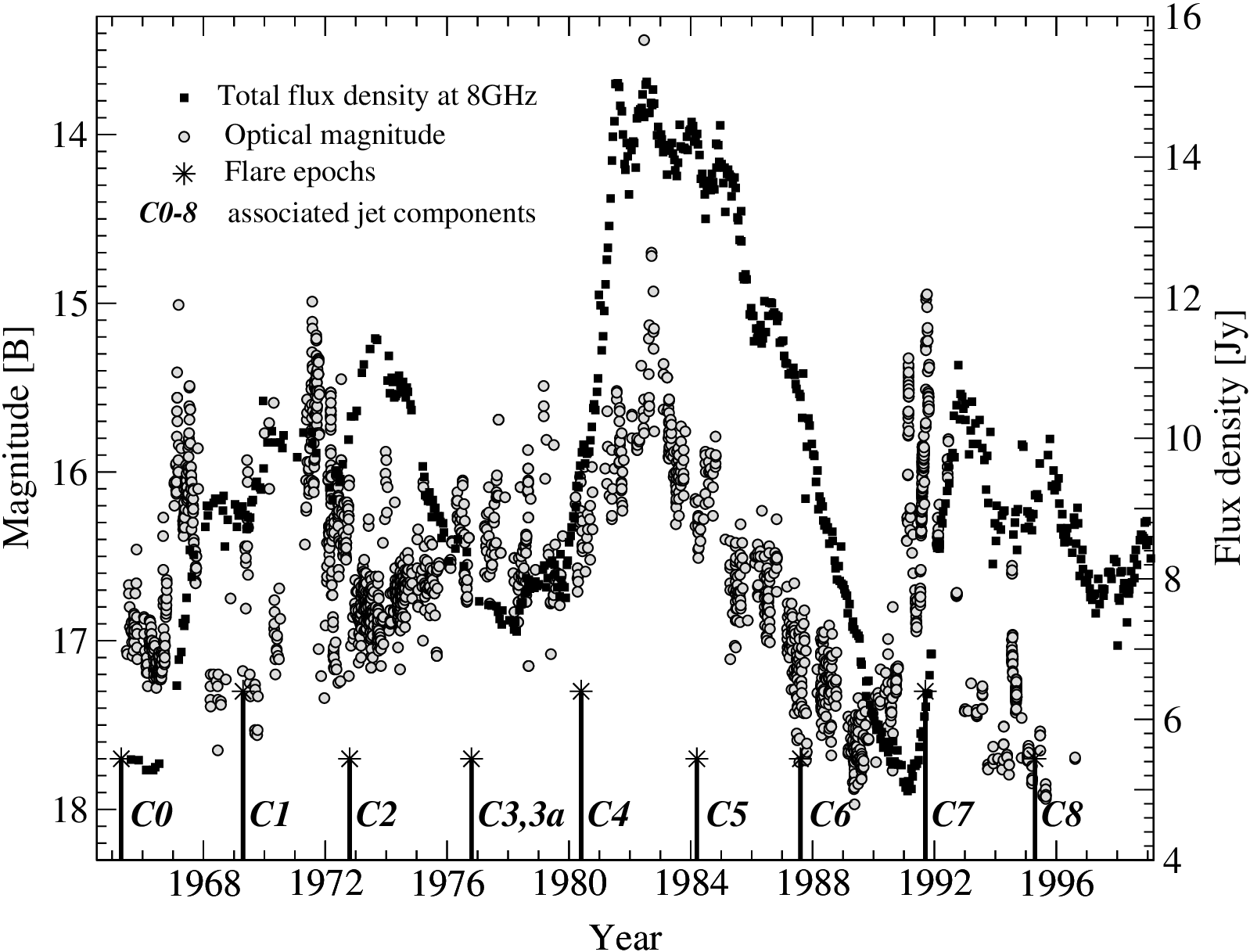}}
\caption{
  Optical and radio variability of \object{3C\,345}. Shown are the epochs of
  spectral flares (\cite{LZ99}) that have resulted in appearances of
  new superluminal components in the jet. The radio data are from the
  Michigan monitoring program ( Aller, Aller \& Hughes
  \cite{aller03}).  The optical data are from Kinman et al.
  (\cite{kin+68}); Smyth \& Wolstencroft (\cite{sw70}); L\"u
  (\cite{lu72}); McGimsey et al. (\cite{mcg+75}); Pollock et al.
  (\cite{pol+79}); Angione et al. (\cite{ang+81}); Kidger
  (\cite{kid88}) Webb et al. (\cite{web+88}); Kidger \& de Diego
  (\cite{kd90}); Vio et al (\cite{vio+91}); Schramm et al.
  (\cite{sch+93}); Babadzhanyants, Belokon' \& Gamm (\cite{bbg95});
  Belokon' \& Babadzhanyants (\cite{bb99}).  All optical data
  have been converted into the B magnitude scale.}
\label{fg:lcurve}
\end{figure*}

\section{The quasar 3C\,345}
\label{sc:sect3}

The 16$^{m}$ quasar \object{3C\,345} exhibits remarkable structural and
emission variability on parsec scales around a compact unresolved
radio core.  The total radio flux density of \object{3C\,345} has been
monitored at 5, 8, and 15\,GHz (Aller, Aller \& Hughes
\cite{aller03}), and at 22 and 37\,GHz (Ter\"asranta et al.
\cite{ter+98}). The source has also been monitored with the Green Bank
Interferometer at 2.7 and 8.1\,GHz (Waltman et al. \cite{wal+91}).
The observed variability of the optical emission is possibly
quasi-periodic with a period of $\approx$1560 days (Babadzhanyants \&
Belokon' 1984; Kidger 1989), although it has been suggested that the
light curve may originate from a non-linear and non-stationary
stochastic process (Vio et al. \cite{vio+91}). A strong optical flare
observed in \object{3C\,345} in 1991--92 (Schramm et al. \cite{sch+93}; Babadzhanyants,
Belokon' \& Gamm \cite{bbg95}) was connected with the appearance of a new
superluminal jet component, C7 (\cite{L96}). \cite{LZ99} report possible 3.5--4 year
quasi-periodicity manifested by spectral flares of the radio emission
and appearances of new jet components.  Historical optical and 8\,GHz
radio lightcurves are shown in Fig.~\ref{fg:lcurve} together with the
epochs of the spectral flares and component ejections.

The evolution of the parsec-scale jet of \object{3C\,345} has been studied
extensively with the VLBI: see Unwin et al. (\cite{unw+83}); Biretta,
Moore \& Cohen (\cite{bmc86}); Zensus, Cohen \& Unwin (1995, hereafter
\cite{ZCU95}); \cite{L96}; Ros, Zensus \& Lobanov (2000, hereafter
\cite{R00}); Klare (\cite{kla03}).  The relativistic jet model
(Blandford \& Rees \cite{br78}) has been applied to explain the
emission and kinematic changes observed in the parsec-scale
structures. Using the X-ray data to constrain the jet kinematics,
\cite{ZCU95} and Unwin et al. (\cite{unw+94}) derive physical
conditions in the jet from a model that combines the inhomogeneous jet
model of K\"{o}nigl (\cite{kon81}) for the core with homogeneous
synchrotron spheres for the moving jet components (Cohen
\cite{coh85}). The evolution of the core flux density is well
represented by a sequence of flare-type events developing in a
partially opaque, quasi-steady jet (\cite{LZ99}).  Spectral properties
of the parsec-scale emission have been studied in several works
(\cite{L96}; Lobanov, Carrara, \& Zensus \cite{lcz97}; Lobanov
\cite{lob98b}).  Unwin et al. (\cite{unw+97}) have analyzed a
correlation between the X-ray variability and parsec-scale radio
structure of \object{3C\,345}.

\begin{figure*}
\centerline{
\includegraphics[width=0.95\textwidth, bb =0 0 585 530, clip=true]
{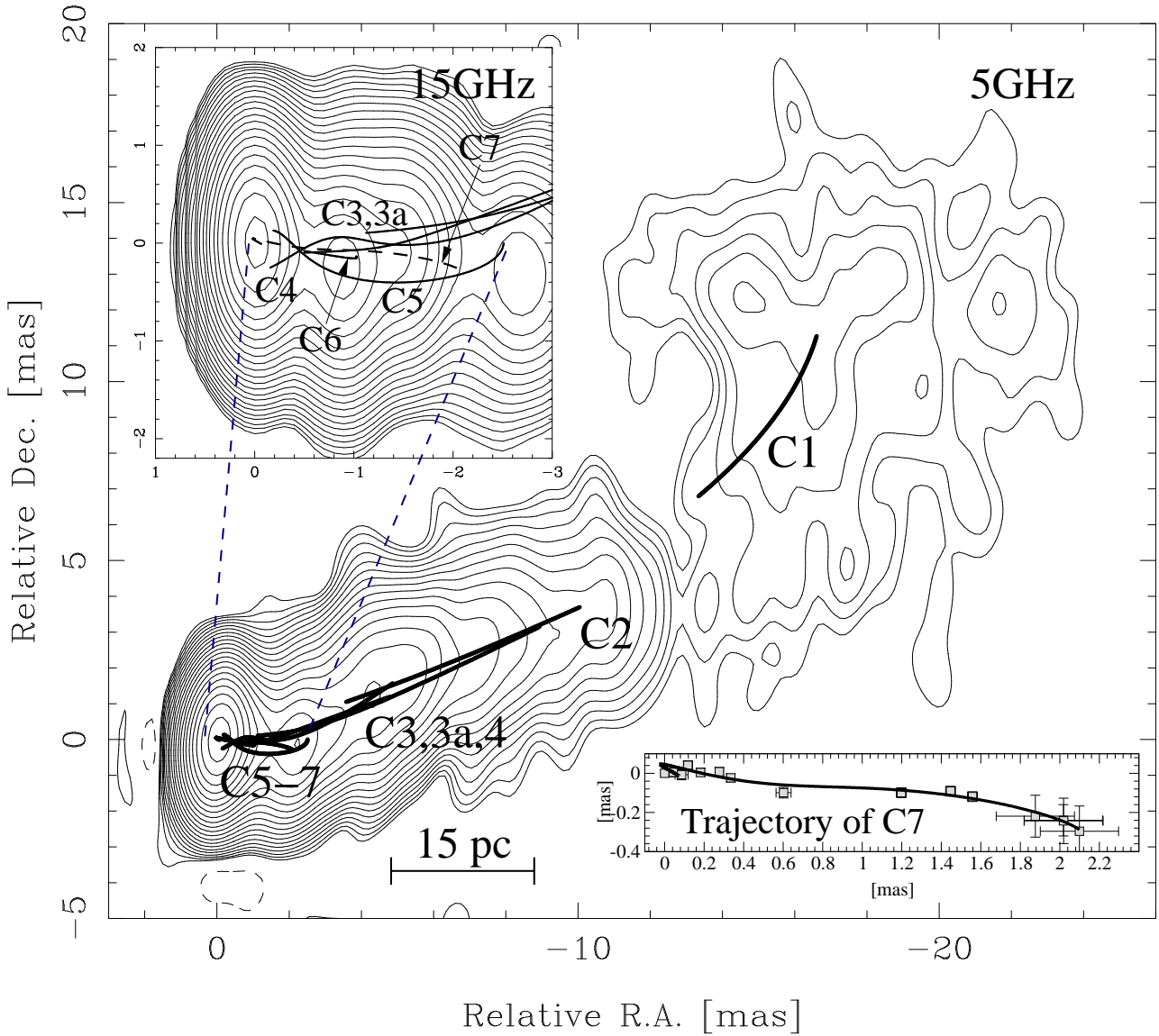}}
\caption{Parsec-scale jet in \object{3C\,345} at 5 and 15\,GHz. Contours are
drawn at -1,1,$\sqrt{2}$,2,... of the lowest contour (2.5\,mJy/beam,
in both images). Curved lines mark the trajectories of superluminal
features C1--C7 identified in the jet. The
component trajectories are polynomial fits to the position
measurements from the 1979--1998 VLBI monitoring data (\cite{ZCU95}; \cite{L96};
\cite{R00}). The positions of C7 measured at
22\,GHz are shown in the inset.}
\label{fg:3c345jet}
\end{figure*}

\begin{table}[ht]
\label{tb:c7}
\caption{Positions and flux densities of C7 at 22\,GHz}
\begin{center}
\begin{tabular}{lrrr}\hline\hline
\multicolumn{1}{c}{$t_\mathrm{ obs}$} & \multicolumn{1}{c}{$S$} & \multicolumn{1}{c}{$\Delta x$} & \multicolumn{1}{c}{$\Delta y$} \\
\multicolumn{1}{c}{$[{\mathrm year}]$} & \multicolumn{1}{c}{$[{\mathrm Jy}]$} & \multicolumn{1}{c}{$[{\mathrm mas}]$} & \multicolumn{1}{c}{$[{\mathrm mas}]$} \\ \hline
1989.25$^{\dag}$ & $0.19\pm0.02$ & $0.04\pm0.02$ & $ 0.01\pm0.01$  \\ 
1991.46 & $0.80\pm0.04$ & $0.09\pm0.05$ & $-0.01\pm0.01$  \\ 
1991.86 & $3.29\pm0.49$ & $0.12\pm0.01$ & $ 0.03\pm0.01$  \\ 
1992.45 & $6.20\pm0.36$ & $0.18\pm0.01$ & $ 0.00\pm0.01$  \\ 
1992.86 & $5.49\pm0.30$ & $0.28\pm0.01$ & $ 0.01\pm0.01$  \\ 
1993.13 & $3.60\pm0.10$ & $0.34\pm0.01$ & $-0.02\pm0.01$  \\ 
1993.72 & $2.23\pm0.12$ & $0.60\pm0.04$ & $-0.10\pm0.02$  \\ 
1994.45 & $1.04\pm0.10$ & $0.90\pm0.09$ & $-0.12\pm0.08$  \\ 
1995.85 & $1.21\pm0.06$ & $1.20\pm0.08$ & $-0.10\pm0.08$  \\ 
1996.42 & $0.88\pm0.05$ & $1.36\pm0.08$ & $-0.07\pm0.07$  \\ 
1996.82 & $0.69\pm0.04$ & $1.55\pm0.08$ & $-0.08\pm0.12$  \\ 
1997.40 & $0.64\pm0.10$ & $1.87\pm0.10$ & $-0.22\pm0.13$  \\ 
1997.62 & $0.59\pm0.07$ & $2.02\pm0.10$ & $-0.24\pm0.12$  \\ 
1997.90 & $0.47\pm0.07$ & $2.10\pm0.10$ & $-0.30\pm0.11$  \\ 
1998.02 & $0.47\pm0.08$ & $2.02\pm0.25$ & $-0.24\pm0.14$  \\ 
1998.99 & $0.50\pm0.10$ & $2.63\pm0.20$ & $-0.15\pm0.14$  \\ 
1999.12 & $0.28\pm0.10$ & $2.69\pm0.20$ & $-0.36\pm0.14$  \\ 
1999.64 & $0.17\pm0.20$ & $2.87\pm0.20$ & $-0.34\pm0.15$  \\ 
\hline
\end{tabular}
\end{center}
Note: $\dag$ -- detection is likely to be confused due to blending
between the core and the preceeding jet component C6. 
References:~1989.25--1993.72 (\cite{L96}); 1994.45 (Lepp\"anen, Zensus
\& Diamond \cite{lzd95}); 1996.42--1997.40
(\cite{R00});  1997.62--1999.64 (Klare \cite{kla03}).
\end{table}

\subsection{Compact jet in 3C\,345}

Figure~\ref{fg:3c345jet} shows VLBI images of the jet in \object{3C\,345} at 5
and 15\,GHz, with the trajectories of eight superluminal components
identified and monitored in the jet in 1979--1999.  The morphology of
\object{3C\,345} is typical for a core dominated source: a bright and compact
core responsible for 50--70\% of the total emission, and a compact
jet that contains several enhanced emission regions (jet components)
moving at apparent speeds of up to $\beta_\mathrm{{app}}\approx20h^{-1}$.

The compact jet in \object{3C\,345} extends over $\sim25$\,mas, which
corresponds to a projected distance of
$\approx$95$\,(h\,\sin\theta_\mathrm{ jet})^{-1}$\,pc
($\approx$800\,$h^{-1}$\,pc, assuming the most likely viewing angle of
the jet $\theta_\mathrm{ jet}=7\degr$) . The jet has two twists at the
separations of $\approx$1 and $\approx$5\,mas. The twists appear to
retain their positions in the jet over the entire duration of the VLBI
monitoring.  Such stability of the twists suggests that they are
likely to reflect the three-dimensional structure of the jet. The
unchanging positions of the twists are also consistent with the
similarities of the apparent trajectories of the jet components at separations
$\ge$1\,mas (Fig.~\ref{fg:3c345jet}).  The two-dimensional
trajectories observed in several jet components follow similar,
overlapping tracks, and outline a continuous picture of a curved jet.
The separation of 20 mas from the core can mark a transition point
from compact jet to large-scale jet.  At this separation, the jet
finally turns to the direction $\phi\approx-40\degr$ observed in the
large-scale jet of \object{3C\,345} (Kolgaard, Wardle \& Roberts \cite{kwr89}).

The measured positions of the component C7 (Fig.~\ref{fg:3c345jet})
provide an accurate account of the jet kinematics within $\sim$2\,mas
distance from the core. The large amount and good quality of the data
at 22\,GHz are sufficient for determining the component trajectory
without using measurements made at lower frequencies. The total of 18
VLBI observations of \object{3C\,345} at 22\,GHz are combined for the
epochs between 1989 and 1998 (\cite{L96}; \cite{R00}; Klare
\cite{kla03}).  For several of these observations, new model fits have
been made for the purpose of obtaining consistent error estimates for
the positions and flux densities of C7.  Table~\ref{tb:c7} presents
the resulting database of the positions and flux densities of C7 at
22\,GHz.

C7 was first reliably detected in 1991.5 at 22\,GHz
at a core separation of $\Theta_{1991.5} \approx 100\,\mu$as (\cite{L96})
Reports of earlier detections in 1989.2 at 100\,GHz (B{\aa}{\aa}th et
al. \cite{baa+92}) and 22\,GHz (\cite{L96}) are inconclusive because
of the likely confusion and blending with the nearby component C6 and
the core itself. 

Following the procedure established in \cite{ZCU95} and \cite{L96},
polynomial functions are fitted to the component coordinate offsets
$x(t)$ and $y(t)$ from the core position which is fixed at the point
(0,0). From the polynomial fits, the proper motion $\mu (t) =
[({\mathrm d} x/{\mathrm d} t)^2 + ({\mathrm d} y/{\mathrm d}
t)^2]^{1/2}$, apparent speed $\beta_\mathrm{ app}(t) = \mu(t)
D_\mathrm{ L} (1+z)^{-1}$ and apparent distance ${\Theta}_{app}(t) =
\int \mu(t^{\prime}) dt^{\prime}$ travelled by the component along its
two-dimensional path can be recovered.  For the purpose of comparison
to the previous work of L96 and Lobanov \& Zensus (\cite{lz96}), the
detection reported for the epoch 1989.25 is included in the polynomial
fitting, but all results for the epochs
before 1991 should be viewed as only tentative.

\begin{figure*}
\centerline{
\includegraphics[width=0.98\textwidth, bb =5 20 767 578, clip=true]
{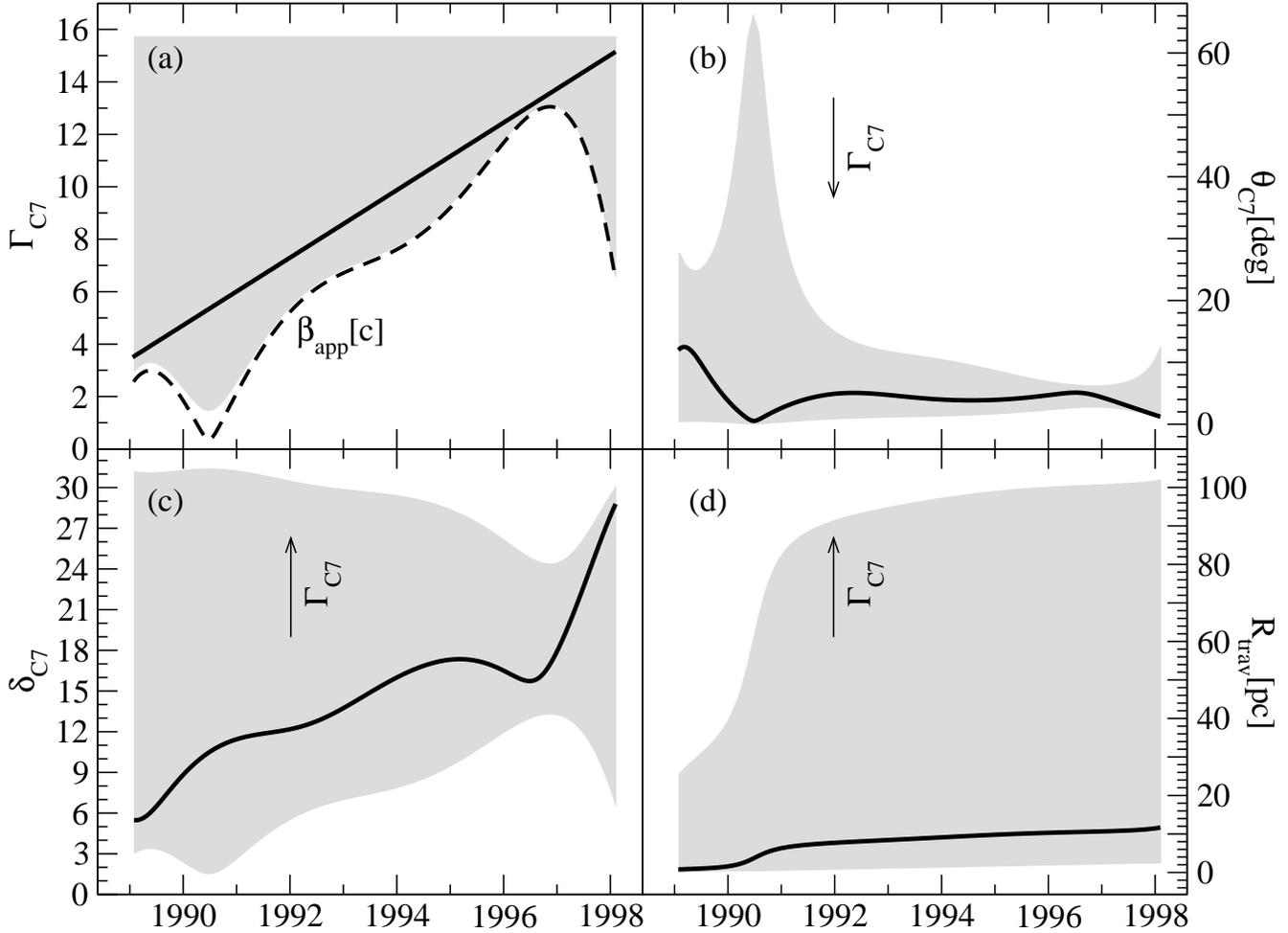}}
\caption{Kinematic properties of the jet determined from the observed 
  trajectory of C7. In all panels, the solid line represents a model
  with $\Gamma_\mathrm{ C7}$ increasing linearly with time from
  $\Gamma_\mathrm{ C7}=2.2$ in 1990 to $\Gamma_\mathrm{ C7}=10.6$ in 1998.  The
  lower boundary of the shaded area corresponds to $\Gamma_\mathrm{ C7}=
  \Gamma_\mathrm{ min} = (\beta_\mathrm{ app}^2+1)^{1/2}$, the upper boundary
  is drawn for $\Gamma_\mathrm{ C7}=11$.  Arrows indicate the direction of
  increasing $\Gamma_\mathrm{ C7}$ from $\Gamma_\mathrm{ min}$ to 11.
  Individual panels show: (a) Measured apparent speed $\beta_\mathrm{
    app}$ (dashed line) and evolution of the Lorentz factor $\Gamma_\mathrm{
    j}$.  (b) Angle between the velocity vector of C7 and the line of
  sight. (c) Doppler factor.  (d) Distance travelled by C7 in the rest
  frame of the jet. The derived $R_\mathrm{ trav} > \beta_\mathrm{ C7} \Delta
  t_\mathrm{ obs}$ because of the time contraction in the observer's frame
  of reference.}
\label{fg:c7kin}
\end{figure*}

\subsection{Kinematics of the jet component C7}

Kinematic properties of C7 are presented in Fig.~\ref{fg:c7kin}.  The
apparent speed $\beta_\mathrm{ app}$ of C7 increased from $\approx
2\,c$ in 1989 to $\beta_\mathrm{ max} \approx 13\,c$ in 1997 (the
decrease of $\beta_\mathrm{ app}$ after 1997 is an artifact of the
polynomial fit).  The component reached the minimum apparent speed of
$\beta_\mathrm{ min} \approx 0.5\,c$ in 1990.5. Such large changes of
$\beta_\mathrm{ app}$ indicate that C7 has moved along a substantially
curved three-dimensional path, with the bulk Lorentz factor most
likely changing with time.  If the Lorentz factor is constant, then
$\Gamma_\mathrm{ C7}>13$ (to satisfy the observed $\beta_\mathrm{
app}$). If $\Gamma_{\rm mathrm C7}$ varies, the lower boundary of the
variations is given by the minimum Lorentz factor $\Gamma_\mathrm{
min} = [\beta_\mathrm{ app}^2(t) + 1]^{1/2}$, which describes the
motion with the least kinetic power.  In Fig.~\ref{fg:c7kin},
$\Gamma_\mathrm{ max}=15.5$ is set, and the kinematic parameters of C7
are described for all allowed variations of $\Gamma_\mathrm{ C7}(t)$
($\Gamma_\mathrm{ min}(t) \le \Gamma_\mathrm{ C7}(t) \le
\Gamma_\mathrm{ max}$).  As seen in Fig.~\ref{fg:c7kin}, neither of
the two boundaries provides a satisfactory description of the
component kinematics.  The case $\Gamma_\mathrm{ C7} = \Gamma_\mathrm{
max}$ results in uncomfortably small $\theta_\mathrm{ C7}(t)$ and
disproportionally large travelled distance $R_\mathrm{ trav}(t)$.
With $\Gamma_\mathrm{ C7} = \Gamma_\mathrm{ min}(t)$, the resulting
variations of $\theta_\mathrm{ C7}$ exceed $50\degr$,
which is difficult to explain.  A plausible evolution of the rest
frame speed of C7 can be described by a linear increase of the Lorentz
factor from $\Gamma_\mathrm{ C7}=2.2$ in 1989 to $\Gamma_\mathrm{ C7}
= 15$ in 1998.  In this case, none of the kinematic parameters change
drastically in the course of the component evolution:
$\theta_\mathrm{ C7}$ varies within $\approx 10\degr$, and $R_\mathrm{
trav} \le 10$\,pc. This indicates that the motion of C7 is likely to be
accelerated during the initial stages of its evolution, at distances
of $\le 2$\,mas from the core. At such distances, acceleration of the
component rest frame speeds has also been reported in several other
jet components in \object{3C\,345} (Lobanov \& Zensus \cite{lz96},
\cite{LZ99}). Consequently, this must be a general property of the jet
plasma on these spatial scales, and it should be taken into account in
the modelling of the jet kinematics.

\subsection{Long term trends in the jet: evidence for precession}

The observed consistency of the two-dimensional component tracks
plotted in Fig.~\ref{fg:3c345jet} suggests that all jet
components in \object{3C\,345} travel along similar trajectories.
There is however a substantial amount of evidence for global evolution
of the entire jet. In Fig.~\ref{fg:slopes}, the apparent
accelerations, ${\mathrm d}\mu/{\mathrm d}\Theta_\mathrm{ app}$, of the components are
plotted against the respective component origin epochs taken from
\cite{L96}.  The apparent accelerations show a steady increase,
following the order of component succession, which is a strong
indication for long-term changes in the jet.  While it is conceivable
that the younger components are intrinsically faster, this would
require the presence of a non-stationary process affecting
significantly the particle acceleration in the nuclear regions of
\object{3C\,345}.  The more likely explanation of the observed
long-term trend of the component accelerations is precession of the
jet axis.  In this case, variations of the component ejection angle
alone are sufficient to explain the observed trend: the components
ejected at angles closer to $\theta_\mathrm{ fast} = 1/\Gamma_\mathrm{
  b}$ would have higher ${\mathrm d}\mu/{\mathrm d}\Theta_\mathrm{
  app}$.  In the course of the precession, the ejection angle would
change steadily, and the measured ${\mathrm d}\mu/{\mathrm
  d}\Theta_\mathrm{ app}$ should show sinusoidal variations.  The
rather large errors in the measured ${\mathrm d}\mu/{\mathrm
  d}\Theta_\mathrm{ app}$ in Fig.~\ref{fg:slopes} allow only a very
rough estimate of the precession period $T_\mathrm{ prec}$, which should be in
the range of 850--3670\,years, to be made.

\begin{figure}
\centerline{
\includegraphics[width=0.45\textwidth, bb =0 0 758 579, clip=true]
{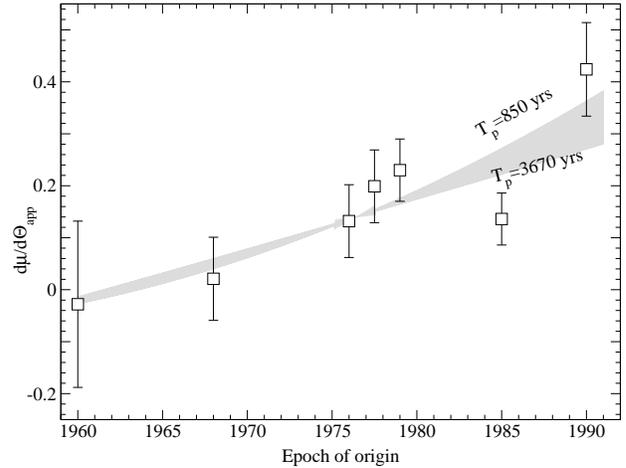}}
\caption{Apparent acceleration ${\mathrm d}\mu/{\mathrm d}\Theta_\mathrm{ app}$ 
  of the component proper motions, plotted versus the component epochs
  of origin. The origin epochs are obtained by back-extrapolating the
  observed trajectories of the jet components (\cite{L96}). The shaded area
  cover the range of precession periods which would reproduce the
  observed trend of ${\mathrm d}\mu/{\mathrm d}\Theta_\mathrm{ app}$.}
\label{fg:slopes}
\end{figure}

Periodic changes are also evident in the variations of the jet
position angle measured at 22\,GHz for successive jet components as
they reach a $\approx$0.5\,mas separation from the core. The resulting
position angles are plotted in Fig.~\ref{fg:pa-change}. The position
angle varies with a period of $P_\mathrm{ short} \approx$9.5 years and
an amplitude of $A_\mathrm{ short} \approx$20$\degr$. A longer term
trend is suggested by the slope in the linear regression of the
position angle data, which implies a rate of change in the position
angle of $\approx$0.4$\degr$/year. The time coverage of the data is
too short to provide an accurate estimate of the period of the long
term variations. A formal fit constrains the period to within a range
of 300--3000 years, with a mean value of $P_\mathrm{ long} \approx
1600$ years and an amplitude of $A_\mathrm{ long}\approx 10\degr$. The
shaded area in Fig.~\ref{fg:pa-change} represents a combined fit to
the data by the two periods. The estimate of $P_\mathrm{ long}$
obtained from the change in position angle falls within the range of
periods implied by ${\mathrm d}\mu/{\mathrm d}\Theta_\mathrm{ app}$,
and it can therefore be associated with the precession. The nature of
the short-term changes of the jet position is unclear, but they could
be related to the rotation of the accretion disc or the jet itself or
to the orbital motion in the BBH system.

\begin{figure}
\centerline{
\includegraphics[width=0.45\textwidth, bb = 24 44 755 508, clip=true]
{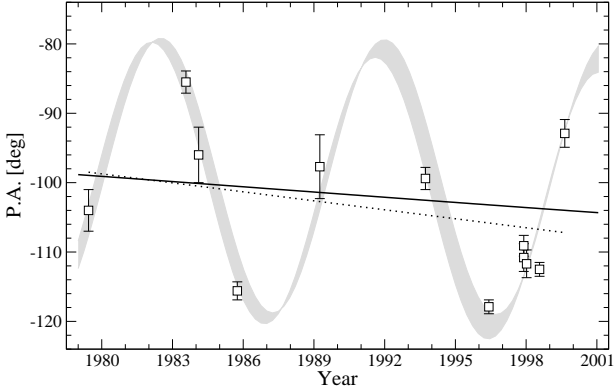}}
\caption{Position angle of different jet components measured at
  22\,GHz at 0.5\,mas separation from the core. Short-term variations, with a
  period of $\approx$9.5 years, are evident. A linear regression
  (dotted line) suggests the presence of a long-term trend, with a
  annual change rate of 0.4$\pm$0.3$\degr$/year. The shaded area shows
  the best fit by the two periodical processes combined, with the
  estimated periods $P_\mathrm{ short} = 9.5\pm0.1$\,years and $P_\mathrm{
    long} = 1600\pm1300$\,years. The solid line represents the trend in the
  position angle due to the long-term periodicity.}
\label{fg:pa-change}
\end{figure}

\section{Application of the BBH model}
\label{sc:sect4}

The BBH model applies a single model framework to explain the
emission and kinematic properties of the jet in the optical and radio
regimes.  The evolution of superluminal features observed in the radio
jet is explained by a precession of the accretion disk around the
primary black hole M$_1$, and the characteristic shape of the optical
light curve is produced by the orbital motion of the black hole M$_1$.

The different characteristic timescales and amplitudes of changes in
the radio and optical regimes imply a complex composition of the
emitting material. The optical emission of \object{3C\,345} during the strong
flare in 1990--93 varied by $\approx 2.5$\,mag, with a typical
timescale of $\sim$0.6\,yr for individual sub-flare events (see
Fig.~\ref{fg:opt_fit}).  The radio emission of the jet component C7
associated with that flare showed only a single rise and decay
event, changing by a factor of $\sim$15 on a timescale of $\sim$6
years.  This difference in the behavior of the optical and radio
variability can be reconciled by assuming that the relativistic
$e^{\pm}$ plasma responsible for the optical emission is injected into
the beam during a time $\Delta\tau_\mathrm{ opt}$ which is much
shorter than the injection timescale $\Delta\tau_\mathrm{ rad}$ for the
plasma responsible for the radio emission.  For C7, the plasma
responsible for the optical emission is modelled by a point-like 
component and the plasma responsible for the radio
emission is modelled by a temporally extended
component, so that $\Delta\tau_\mathrm{ opt} <
0.01\Delta\tau_\mathrm{ rad}$. In the BBH frame, the
radio component emits over a period of $\approx 1.9$\,yr, with
component ejections occurring approximately every 4 years.  A general
sketch of the composition of a flare is shown in
Fig.~\ref{fg:flaresketch}.

The application of the model consists of two parts. In the first
part, the observed trajectory and flux density of the jet component C7
are reconstructed, using the precession of the accretion disk around
the black hole M$_1$, and solving formally the problem defined by
Eq.~(\ref{eq:beta_quad}).  In the second part, the optical light
curve of \object{3C\,345} is calculated from the orbital properties of the
BBH system provided by the solution of Eq.~(\ref{eq:tz_final}). 

\begin{figure}[ht]
\centerline{
\includegraphics[width=0.48\textwidth, bb =0 0 509 345, clip=true]
{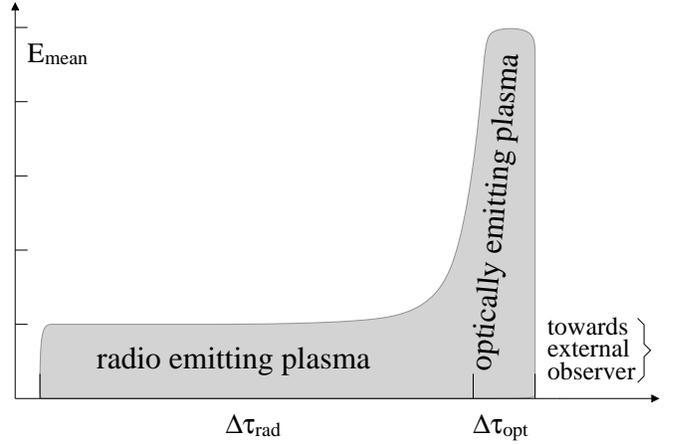}}
\caption{Composition of a nuclear flare. The flare begins with an
injection of highly energetic particles emitting optical synchrotron
radiation. The timescale of the injection $\Delta\tau_\mathrm{ opt}$
is comparable to the synchrotron loss time of the optically emitting
plasma.  Rapid energy losses and injection of radio emitting plasma on
the timescale $\Delta\tau_\mathrm{ rad}$ lead to progressively decreasing
mean energy of the plasma and subsequent propagation of the flare to
radio wavelengths. The optically emitting plasma is represented in the
model by a
point-like component injected into the beam, while the radio 
emitting plasma is temporally resolved, with the
resulting $\Delta\tau_\mathrm{ opt} <
0.01\Delta\tau_\mathrm{ rad}$
}
\label{fg:flaresketch}
\end{figure}

\subsection{Precession and geometry}

Basic geometrical parameters of the system are recovered by fitting
the precession model to the observed two-dimensional path of C7. These include
\begin{description}
\item[]~$\iota$~--~viewing angle of the jet axis (Fig.~\ref{fg:scheme});
\item[]~$\Psi_\mathrm{ proj}$~--~position angle of the jet axis in the
  plane of the sky;
\item[]~$T_\mathrm{ p}$~--~precession period;
\item[]~$\Omega_\mathrm{ p}$~--~opening half-angle of the precession cone;
\item[]~$\phi_\mathrm{ p}$~--~initial phase of the precession;
\item[]~$R_0$~--~maximum amplitude of the perturbation;
\item[]~$t_\mathrm{ beam}$~--~characteristic dampening time of the perturbed
  beam.
\end{description}

It should be stressed that none of the parameters listed above depend
on the velocity of the moving component. However, the shape of the
path depends on the Alfv\'en speed $V_\mathrm{ A}$ of the magnetic field
perturbation. This speed is determined by fitting simultaneously the
geometry of the component path and the orbital motion in the binary
system.

\subsubsection{Velocity of the jet component}

The velocity of the jet component is constrained by fitting the coordinate
offsets, $x(t)$ and $y(t)$, of the component from the position of the
radio core. In
\object{3C\,345}, the apparent motion of all jet components is non-linear and
shows accelerations and decelerations (\cite{ZCU95}; \cite{L96}; \cite{LZ99}). To account for
this, a variable Lorentz factor must be introduced, using
Eq.~(\ref{eq:Gamma_t}). With a variable Lorentz factor, the
$x(t)$ and $y(t)$ offsets are fitted simultaneously, and a
three-dimensional trajectory of the component is recovered.
This procedure yields the following parameters:

\begin{description}
\item[]~$\Gamma_\mathrm{ min}$~--~initial Lorentz factor of the component;
\item[]~$\Gamma_\mathrm{ max}$~--~final Lorentz factor of the component,
  after the acceleration;
\item[]~$t_\mathrm{ acc}$~--~characteristic acceleration time.
\end{description}

\subsection{Flux density changes in the jet component}

In addition to the geometrical and kinematic factors, the flux density
evolution in the optical and radio regimes is characterized by the
synchrotron energy loss timescales $\Delta\tau_\mathrm{ opt}$ and
$\Delta\tau_\mathrm{ rad}$ and by the duration 
$\Delta\tau_\mathrm{ thick}$ during which the relativistic plasma
responsible for the flare remains optically thick in the radio
regime.  The radio lightcurve of C7 constrains uniquely the viewing
angle of the jet. The information contained in the radio lightcurve is
fundamental for determining the duration of the radio emitting stage,
$\Delta\tau_\mathrm{ rad}$, of the flare and ensuring the consistency
between the precession fit and the BBH fit.

\subsection{BBH system}

The main effect of the BBH system is the orbital motion of both black
holes around the center of mass of the system. This motion explains
the observed optical light curve, the fine structure of the parsec-scale jet,
and peculiarities of component motions around the general trajectory
determined by the precession.  The fit by the BBH system yields the
following parameters:

\begin{description}
\item[]~$M_{1}$~--~mass of the primary black hole containing the
  accretion disk;
\item[]~$M_{2}$~--~mass of the secondary black hole;
\item[]~$T_\mathrm{ o}$~--~orbital period in the BBH system;
\item[]~$e$~--~eccentricity of the orbit;
\item[]~$t_\mathrm{ o}$~--~injection epoch of a new jet component;
\item[]~$\phi_\mathrm{ o}$~--~initial orbital phase at the injection epoch.
\end{description}

\subsection{Consistency between the precession and BBH fits}
\label{sc:consistency}

The BBH model, which is obtained by fitting a
point-like component to reproduce the observed optical lightcurve,
must be consistent with the results derived from the precession fit.
In  other words, the evolution of the extended, radio emitting, component 
determined from fitting the BBH model must reproduce the
trajectory and kinematic properties of the radio emitting plasma 
obtained from the precession fit.

Consistency between the fits by the precession and orbital motion is
ensured under the following five conditions which connect the formal
fitting with the specific properties of the two-fluid model:

\begin{description}
  
\item[1.]~The characteristic Lorentz factor obtained from fitting the
  optical light curve by the BBH model must be larger than the
  corresponding Lorentz factor obtained from fitting the trajectory of
  C7 by the precession model. This is simply due to the fact that the
  rest frame trajectory in the BBH model is longer than the trajectory
  in the precession model (see Fig.~\ref{fg:c7_traj1}, implying that
  the component has to move faster in the BBH model to
  reproduce the same observed two-dimensional trajectory.

\item[2.]~The optical component is assumed to be temporarily
  unresolved, but not the radio component. This
  condition is required to explain the differences in the emission
  variations observed in the optical and radio regimes. 

\item[3.]~The ejection epoch of the optical component precedes that of
the radio component. This condition reflects the nuclear flare
structure assumed (Fig.~\ref{fg:flaresketch}).

\item[4.]~For a given mass $M_1$ of the primary black hole, the
combination of the mass of the secondary $M_2$, orbital period $T_\mathrm{
o}$ and acceleration time $t_\mathrm{ acc}$ must reconcile satisfactorily
the rest frame time derived from the precession model with the
rest frame time obtained from the BBH model (see
Fig.~\ref{fg:trest-tobs}).

\item[5.]~A single value of the Alfv\'en speed must reproduce the observed
  shape of the component trajectory obtained from the precession fit and
  satisfy the orbital period obtained from the BBH fit.

\end{description}

These five conditions ensure that the kinematic and flux density
variations of the radio component recovered from the BBH and
precession fits are consistent, and thus the BBH scenario explains
the radio and optical components simultaneously.

\section{Properties of the BBH system in 3C\,345}
\label{sc:sect5}

The best fit parameters of the precessing jet and the BBH system in
\object{3C\,345} determined using the model developed in Sect.~\ref{sc:sect2}
and the fitting method described in Sect.~\ref{sc:sect4} are
presented in Table~\ref{tb:bestfit}.  The procedural steps through
which the fit was obtained are summarized below.

\begin{table*}
\caption{Precession, orbital and flare properties of the BBH system in
  3C\,345}
\label{tb:bestfit}
\begin{center}
\begin{tabular}{llr}\hline\hline
\multicolumn{3}{c}{Geometrical parameters} \\ \hline
Viewing angle of the jet, & $\iota$ & $6.5\degr$ \\
Position angle of the jet in the plane of the sky, & $\Psi_\mathrm{ proj}$ &
$-95\degr$ \\
Opening half-angle of precession cone, & $\Omega_\mathrm{ p}$ & $1.45\degr$ \\
Precession period of accretion disk, &$T_\mathrm{ p}$ & 2570\,yr \\ 
Characteristic dampening (comoving) timescale of perturbation, &$t_\mathrm{ beam}$ & 180\,yr \\
Maximum perturbation amplitude, &$R_0$  & 1\,pc \\\hline
\multicolumn{3}{c}{Kinematic parameters} \\ \hline
Alfv\'en speed, &$V_\mathrm{ A}$ & $0.078\,c$ \\ 
Initial Lorentz factor of radio emitting plasma, &$\Gamma_\mathrm{ min,r}$ & 3.5 \\
Initial Lorentz factor of optically emitting plasma, &$\Gamma_\mathrm{ min,o}$ & 12 \\
Terminal Lorentz factor of jet plasma, &$\Gamma_\mathrm{ max}$ & 20 \\
Plasma acceleration (comoving) timescale, &$t_\mathrm{ acc}$ & 165\,yr \\ \hline
\multicolumn{3}{c}{Parameters of the binary system} \\ \hline
Primary black hole mass, &$M_1$ & $7.1\times 10^8\,{\mathrm M}_{\sun}$ \\
Secondary black hole mass, &$M_2$ & $7.1\times 10^8\,{\mathrm M}_{\sun}$ \\
Orbital period, &$T_\mathrm{ o}$& 480\,yr \\
Orbital separation, &$r_\mathrm{ o}$& 0.33\,pc \\
Orbital eccentricity, & $e$ & $< 0.1$ \\
Characteristic rotational period of accretion disk, &$T_\mathrm{ disk}$&
240\,yr \\ \hline
\multicolumn{3}{c}{Flare parameters and initial conditions} \\ \hline
Plasma injection epoch, & $t_\mathrm{ inj}$ & 1991.05 \\
Initial precession phase, &$\phi_\mathrm{ p}$& 100$\degr$ \\
Initial orbital phase, &$\phi_\mathrm{ o}$& 125$\degr$ \\ 
\hline
\end{tabular}
\end{center}
\end{table*}

\subsection{Geometrical and precession properties}
\label{sc:fit-prec}

The position angle of the jet
axis in the plane of the sky is determined directly from the VLBI images
of \object{3C\,345}, which yields
\[
\Psi_\mathrm{ proj}=-95.5\degr\,.
\]

The injection epoch $t_\mathrm{ o}$ does not have to coincide with the
epoch of first detection of C7, since the temporal and linear scales
for plasma acceleration may be comparable to, or even larger than, the
smallest angular scale at which the component becomes detectable in
VLBI images. 
The most accurate estimate of the epoch of origin of C7 is provided by the 
onset of the radio flux increase (see Fig.~\ref{fg:lcurve}), which yields
\[
t_\mathrm{ o} = 1991.05
\]
This value indeed provides the best fit for both the optical
lightcurve and the evolution of C7. 

The jet viewing angle, $\iota$, opening half-angle of the precession cone
$\Omega_\mathrm{ p}$, and initial precession phase, $\phi_\mathrm{ p}$, are obtained by
fitting the observed two-dimensional path of C7.
This fit constrains uniquely the initial precession phase at $t_\mathrm{ o}$
\[ 
\phi_\mathrm{ p} = 100\degr\,.
\]

For fitting the jet viewing angle, an initial guess is provided by the
best fit values ($\iota = 5$--8$\degr$ obtained from modelling the
kinematic (\cite{ZCU95}), spectral (\cite{LZ99}) and opacity (Lobanov
\cite{lob98a}) 
properties of the compact jet in \object{3C\,345}. A family of
solutions across the $\iota = 5$--$8\degr$ interval is presented in
Table~\ref{tb:sol-iota}, together with corresponding $T_\mathrm{ p}$
and $t_\mathrm{ beam}$ calculated for $V_\mathrm{ A} = 0.078\,c$.
(note that determination of $\iota$ and $\Omega_\mathrm{ p}$ does not
depend on the knowledge of $V_\mathrm{ A}$).  The best fit to the
observed path of C7 is provided by
\[
\iota = 6.5\degr\,, \quad \Omega_\mathrm{ p} = 1.45\degr\,.
\]
At this stage, the maximum perturbation amplitude, $R_0$, and the
timescale, $t_\mathrm{ beam}$ of damping the perturbed beam can be
constrained. An initial guess for $R_0$ (0.5--0.8\,pc) is
provided by the largest deviation of the positions of C7 from a
straight line defined by $\Psi_\mathrm{ proj}$. 
The best fit for the observed path of C7 is obtained with 
\[
R_0 = 1.0\,{\mathrm pc}\,, \quad t_\mathrm{ beam} = 180\,{\mathrm yr}\,.
\]

\begin{table}
\caption{Precession solutions for $\iota = 5$--$8\degr$}
\label{tb:sol-iota}
\begin{center}
\begin{tabular}{c|ccccc}\hline\hline
Parameter & \multicolumn{5}{c}{Value} \\\hline
$\iota$ $[\degr]$ & 5 & 6 & {\bf 6.5} & 7 & 8 \\
$\Omega_\mathrm{ p}$ $[\degr]$ & 1.3 & 1.4 & {\bf 1.45} & 1.5 & 1.6 \\
$t_\mathrm{ beam}$ $[{\mathrm yr}]$ & 220 & 200 & {\bf 180} & 165 & 150 \\
$T_\mathrm{ p}$ $[{\mathrm yr}]$ & 3320 & 2780 & {\bf 2570} & 2370 & 2070 \\ \hline
\end{tabular}
\end{center}
Note: Bold face denotes the best fit solution. 
$T_\mathrm{ p}$ is calculated with the
best fit value of the Alfv\'en speed $V_\mathrm{ A} = 0.078\,c$. 
\end{table}

The fit for $T_\mathrm{ p}$ depends on the Alfv\'en speed $V_\mathrm{ A}$ in
the jet. For each value $V_\mathrm{ A}$ an appropriate $T_\mathrm{ p}$ can be
found from the precession fit that would reproduce exactly the same
two-dimensional trajectory. This family of solutions is given in
Table~\ref{tb:sol-prec}. The value of $V_\mathrm{ A}$ is constrained
after a similar family of solutions is obtained for the BBH fit to the
optical light curve. A unique solution for $V_\mathrm{ A}$ is then found
by reconciling the values of $T_\mathrm{ p}$ and $T_\mathrm{ o}$ obtained from
the BBH and precession fits.

\begin{table}
\caption{Precession solutions for $V_\mathrm{ A} = 0.05\,c$--$0.15\,c$}
\label{tb:sol-prec}
\begin{center}
\begin{tabular}{c|cccccc}\hline\hline
Parameter & \multicolumn{6}{c}{Value} \\\hline
$V_\mathrm{ a}$ $[c]$ & 0.050 & 0.075 & 0.078 & 0.100 & 0.125 & 0.150 \\
$T_\mathrm{ p}$ $[{\mathrm yr}]$ & 4060 & 2690 & 2570 & 1970 & 1510 & 1230 \\ \hline
\end{tabular}
\end{center}
\end{table}

\subsection{Kinematic fit}

After the two-dimensional path of C7 has been fitted by the geometrical
and precession parameters, the precession model is applied to fit the
trajectory and flux density evolution of C7, which yields $\Gamma_\mathrm{
  min} = 3.5$, $\Gamma_\mathrm{ max} = 20$ and $t_\mathrm{ acc} = 165$\,yr.
These parameters are independent from the choice of $V_\mathrm{ A}$.
It should also be noted that the Lorentz factors derived above are
related to the radio emitting plasma. The optically emitting plasma
may have a different range of Lorentz factors.

\subsection{Consistency between the radio and optical fits}
\label{sc:cond-consist}

In order to ensure consistency between the precession and BBH fits,
and to bootstrap together the fit of the trajectory and kinematics
of C7 and the fit of the optical light curve, two general model classes
are considered for describing the properties of the radio and
optically emitting plasmas, in terms of the minimum (or initial)
Lorentz factors of the two plasmas (these scenarios are based on the
consistency conditions described in Sect.~\ref{sc:consistency}):
\newline
{\em Model 1:}~$\Gamma_\mathrm{ min,\,rad} < \Gamma_\mathrm{ min,\,opt}$,
\newline
{\em Model 2:}~$\Gamma_\mathrm{ min,\,rad} = \Gamma_\mathrm{ min,\,opt}$.
\newline
In both cases, the
radio emitting plasma is injected immediately after the optically
emitting plasma (as illustrated in Fig.~\ref{fg:flaresketch}).

For Model 1, $\Gamma_\mathrm{ min,\,rad} = 4$ and $\Gamma_\mathrm{
  min,\,opt} = 12$ are chosen from the precession fit (note that this
corresponds to injection speeds of $v_\mathrm{ min,\,rad} =
0.9682\,c$ and $v_\mathrm{ min,\,opt} = 0.9965\,c$). For Model 2,
$\Gamma_\mathrm{ min,\,rad} = \Gamma_\mathrm{ min,\,opt} = 12$ is assumed.
Both scenarios fit well the optical light curve, but Model 1
reproduces better the kinematics of C7, particularly at the beginning
of the component evolution. 

Fig.~\ref{fg:trest-tobs} compares the
rest frame time obtained from the precession fit and the BBH fit by
the two models. The injection epoch $t_\mathrm{ inj}= 1991.05$ obtained
from the precession fit is recovered by Model 1, but cannot be
reproduced by Model 2.

\begin{figure}
\centerline{
\includegraphics[width=0.45\textwidth, bb =120 70 680 470, clip=true]
{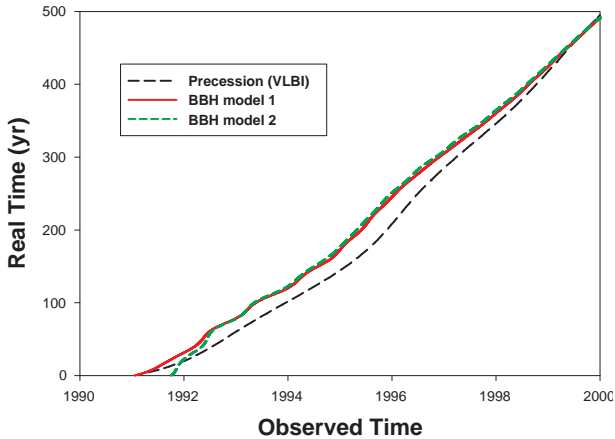}}
\caption{Rest frame time recovered from the precession fit (describing
  the trajectory of the VLBI component) and the BBH Models 1 and 2
  (describing the optical variability) under the consistency
  constraints discussed in Sect.~\ref{sc:consistency}. The injection
  epoch $t_\mathrm{ inj}=1991.05$ (for which $t_\mathrm{ rest}=0$ is required
  by the precession fit) is reproduced by the BBH Model 1 and cannot
  be recovered from the BBH Model 2. The agreement between the
  precession fit and Model 1 ensures that the BBH scenario provides
  a single framework for describing simultaneously the observed
  trajectory and flux density changes of C7 and the observed
  variability of the optical emission of \object{3C\,345}}
\label{fg:trest-tobs}
\end{figure}

For the earliest epochs of the evolution of C7, Model 2 does not
reproduce satisfactorily the separations of C7 from the core
(Fig.~\ref{fg:c7_dist}) and the flux density changes
(Fig.~\ref{fg:c7_flux}).  The discrepancy between the fits by the
precession model and Model 2 is particularly visible in the
speed evolution obtained from the precession and BBH fits
(Fig.~\ref{fg:app-speed}). Model 1 is therefore a
better counterpart to the precession fit described in
Sect.~\ref{sc:fit-prec}.

\begin{figure}
\centerline{
\includegraphics[width=0.45\textwidth, bb =150 85 680 465, clip=true]
{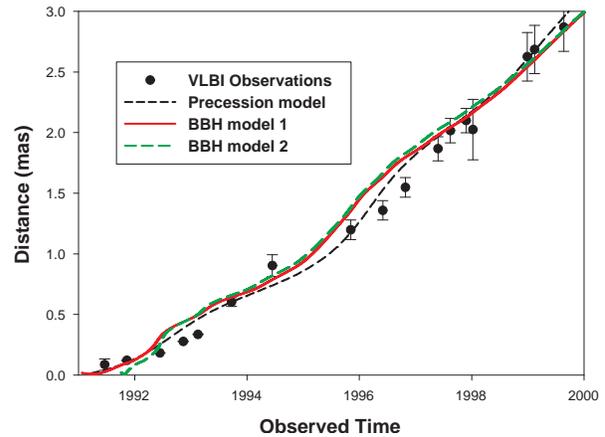}}
\caption{Evolution of the separation of the jet component C7 from the core. 
  The fit by the precession model is consistent with the fit by
  BBH Model 1. BBH Model 2 fails to reproduce the earliest
  position measurements of C7.}
\label{fg:c7_dist}
\end{figure}

\begin{figure}
\centerline{
\includegraphics[width=0.45\textwidth, bb =150 100 660 474, clip=true]
{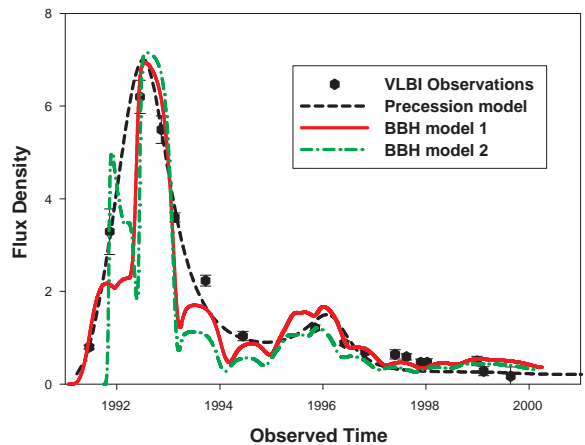}}
\caption{Radio flux density of C7 at 22\,GHz. The fits by
  the precession model and BBH Model 1 are consistent. BBH
  Model 2 cannot reproduce the onset of the radio emission.}
\label{fg:c7_flux}
\end{figure}

\begin{figure}
\centerline{
\includegraphics[width=0.45\textwidth, viewport =150 80 690 465, clip=true]
{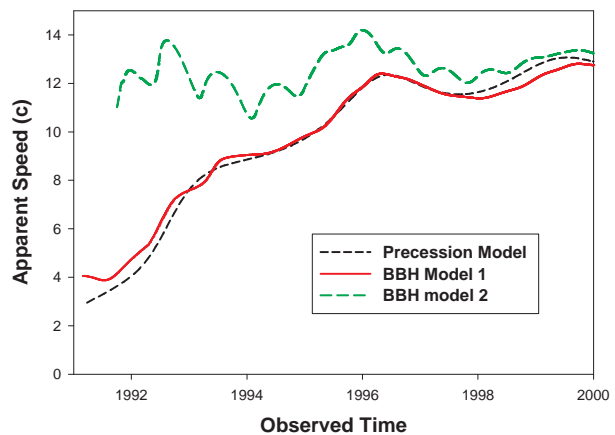}}
\caption{Apparent speed evolution of the jet component C7 recovered
  from the precession and BBH fits. The fit by BBH Model 1 is
  consistent with the fit by the precession, while BBH Model 2
  cannot be reconciled with the precession fit at all.}
\label{fg:app-speed} 
\end{figure}

Figure~\ref{fg:c7_traj2} compares the precession and the BBH Model 1
fits to the observed two-dimensional path of C7. The agreement between
the fits is excellent. The two fits differ by less than 0.02\,mas
(Fig.~\ref{fg:c7_traj1}). This deviation is caused by the orbital motion
of the black hole ejecting the jet, which is not accounted for in the
precession fit.

\begin{figure}
\centerline{
\includegraphics[width=0.45\textwidth, bb =60 258 515 563, clip=true]
{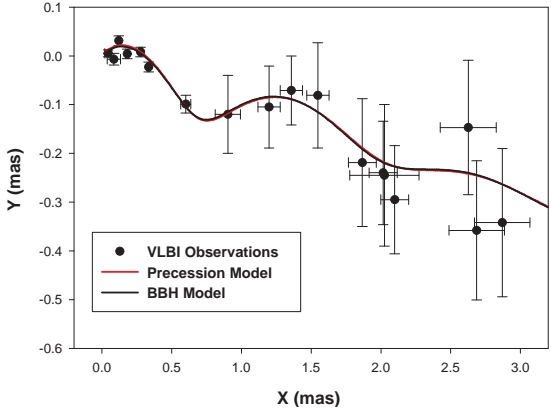}}
\caption{Two-dimensional path of C7.
 The precession fit and the fit by BBH Model 1 are presented.}
\label{fg:c7_traj2}
\end{figure}

\begin{figure}
\centerline{ 
\includegraphics[width=0.45\textwidth, bb =130 70 675 455, clip=true] 
{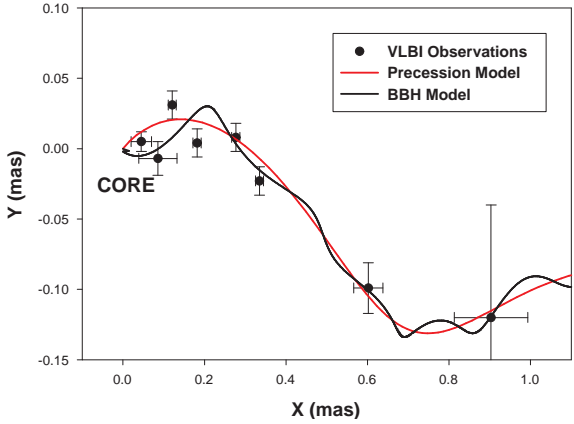}}
\caption{Two-dimensional path of C7 within 1\,mas of the nucleus. 
  The difference between the precession and BBH Model 1 fit is
  due to the orbital motion of the black hole ejecting the jet.}
\label{fg:c7_traj1}
\end{figure}

\subsection{Optical light curve and orbital period}

Fitting the optical light curve constrains the orbital period $T_\mathrm{
  o}$, orbital eccentricity and initial orbital phase $\phi_\mathrm{ o}$
in the BBH system, under the conditions of consistency described in
Sect.~\ref{sc:cond-consist}.  The fit to the optical curve requires
a nearly circular orbit ($e < 0.1$), constrains uniquely the
initial orbital phase
\[
\phi_\mathrm{ o} = 125\degr
\]
and produces a family of solutions for $T_\mathrm{ o}$ depending on the value of
$V_\mathrm{ A}$. These solutions are presented in Table~\ref{tb:sol-tbin} for
the same range of $V_\mathrm{ A}$ as in Table~\ref{tb:sol-prec}. The fit
to the optical light curve is shown in Fig.~\ref{fg:opt_fit}.
It does not depend on the actual value of $T_\mathrm{ o}$ because
$T_\mathrm{ o} V_\mathrm{ A}^{-1} = const$, and thus the same shape of the
lightcurve
can be reproduced by a number of combinations of $V_\mathrm{ A}$ and
$T_\mathrm{ o}$.

\begin{table}
\caption{Orbital period solutions for $V_\mathrm{ A} = 0.05\,c$--$0.15\,c$}
\label{tb:sol-tbin}
\begin{center}
\begin{tabular}{c|cccccc}\hline\hline
Parameter & \multicolumn{6}{c}{Value} \\\hline
$V_\mathrm{ a}$ $[c]$ & 0.050 & 0.075 & 0.078 & 0.100 & 0.125 & 0.150 \\
$T_\mathrm{ 0}$ $[{\mathrm yrs}]$ & 770 & 505 & 480 & 375  & 290 & 250 \\ \hline
\end{tabular}
\end{center}
\end{table}

\begin{figure}
\centerline{
\includegraphics[width=0.45\textwidth, bb =150 100 661 460, clip=true]
{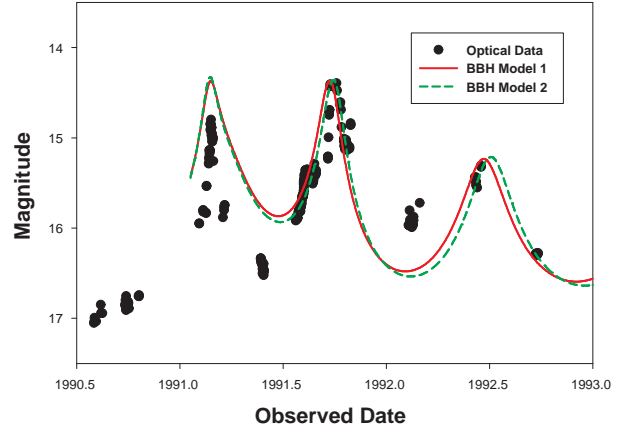}}
\caption{Optical variability in \object{3C\,345} in 1990--93. Solid lines show the
fits by the BBH model. Individual peaks
result from the orbital motion in the BBH system.}
\label{fg:opt_fit}
\end{figure}

\subsection{Determination of the Alfv\'en speed}

In order to reconcile the two families of $V_\mathrm{ A}$-dependent
solutions (the precession solutions in Table~\ref{tb:sol-prec} and
the orbital period solutions in Table~\ref{tb:sol-tbin}), the value of
the Alfv\'en speed must be constrained. The periodic changes of the
position angle of component ejection plotted in
Fig.~\ref{fg:pa-change} can be used for this purpose. The Alfv\'en
wave perturbations, propagating at $V_\mathrm{ A}\ll v_\mathrm{ c}$ cannot
account for such a short period.  The observed period $P_\mathrm{
  short}\approx 9.5$\,yr must therefore reflect perturbations of the
magnetic field caused by the orbital motion in the BBH system. In this
case, $T_\mathrm{ o}  = P_\mathrm{ short} \delta_\mathrm{ c,0.5\,mas}$, where
$\delta_\mathrm{ c,0.5\,mas}$ is the Doppler factor of the beam plasma
measured at the same location (0.5\,mas from the nucleus) at
which the oscillations of the ejection angle are registered. The
kinematic fit yields $\delta_\mathrm{ c,0.5\,mas} = 50$, which results
in the orbital period $P_\mathrm{ o} = 480$\,yr. The corresponding value of the
Alfv\'en speed is then 
\[
V_\mathrm{ A} = 0.078\,c\,.
\]
It should be noted that the Doppler factor $\delta_\mathrm{c,0.5\,mas} = 50$
is similar to the Doppler factors implied by the
observed statistics of superluminal motions near blazar cores (Jorstad et al.
\cite{jor+01a}).

\subsection{Parameters of the binary system}

With the Alfv\'en speed determined, the masses of the black holes can
now be estimated from $T_\mathrm{ o}$ and $T_\mathrm{ p}$. Since the orbital
separation $r_\mathrm{ o} \propto (M_1 + M_2)^{1/3}$, a family of
solutions can be produced that satisfy $T_\mathrm{ o}=480$\,yr, $T_\mathrm{
  p} = 2570$\,yr and reproduce the observed optical lightcurve of
\object{3C\,345} and radio properties of C7 (the shapes of the optical and
radio lightcurves depend on the mass ratio $\mu_{1,2} = M_1/M_2$). The
resulting solutions are presented in Table~\ref{tb:sol-mass} for five
selected values of $M_1$.

In order to identify an acceptable solution for the black hole masses,
stability conditions of the accretion disk must be considered. The
rotational timescale of a precessing accretion disk in a binary system is
given by a characteristic period (Laskar \& Robutel \cite{lr95})
\[
T_\mathrm{ disk} = \frac{4}{3}\left(\frac{M_1}{M_2} + 1\right)
\frac{T_\mathrm{ o}^2}{T_\mathrm{ p}}\,.
\]
The disk in a BBH system will be stable over timescales comparable
with the typical timescales of existence of a powerful radio source in 
AGN, $\tau_\mathrm{ agn}\la
10^8$\,yr, if two conditions are satisfied:

\begin{description}
\item[1.]~$M_1 \ge M_2$, where $M_1$ is the mass of the black hole
  surrounded by the accretion disk.
\item[2.]~$T_\mathrm{ disk} \le T_\mathrm{ o}/2$
\end{description}

These conditions restrict the mass of the primary black hole to a
range $7.1\times 10^8{\mathrm M}_{\sun} \le M_1 \le 7.5\times 10^8{\mathrm
  M}_{\sun}$. The lower limit corresponds to $\mu_{1,2} = 1$, and the
upper limit corresponds to $P_\mathrm{ disk} = P_\mathrm{ o}/2 =
240$\,yr. The acceptable range of $M_1$ is very narrow, and any
value within this range can be adopted. Since $\mu_{1,2}$ does not
change significantly over the acceptable range of $M_1$ ($\mu_{1,2} = 1.02$ for
$M_{1} = 7.5\times 10^8{\mathrm M}_{\sun}$), $\mu_{1,2} = 1$ can be adopted
for the binary black hole mass solution, yielding
\[ 
M_1 = M_2 = 7.1\times 10^8{\mathrm M}_{\sun}
\]
listed in Tables~\ref{tb:bestfit} and \ref{tb:sol-mass}.

\begin{table}
\caption{Black hole mass solutions for $T_\mathrm{ o} = 480$\,yr}
\label{tb:sol-mass}
\begin{center}
\begin{tabular}{c|ccccc}\hline\hline
Parameter & \multicolumn{5}{c}{Value} \\\hline
$M_1$ $[10^8 M_{\sun}]$ & 2.0 & 6.0 & 7.1 & 20.0 & 60.0 \\
$M_2$ $[10^8 M_{\sun}]$ & 4.0 & 6.6 & 7.1 & 12.3 & 23.1 \\
$r_\mathrm{ o}$ $[{\mathrm pc}]$ & 0.25 & 0.32 & 0.33 & 0.44 & 0.60 \\ 
$T_\mathrm{ disk}$ $[{\mathrm yrs}]$ & 179 & 229 & 239 & 314 & 430 \\ \hline
\end{tabular}
\end{center}
Notes: $r_\mathrm{ o}$~--~orbital separation between the black holes,
$T_\mathrm{ disk}$~--~characteristic rotational period of the accretion
disk around the primary black hole.
\end{table}

\begin{figure}
\centerline{
\includegraphics[width=0.45\textwidth, bb = 13 22 734 522, clip=true]
{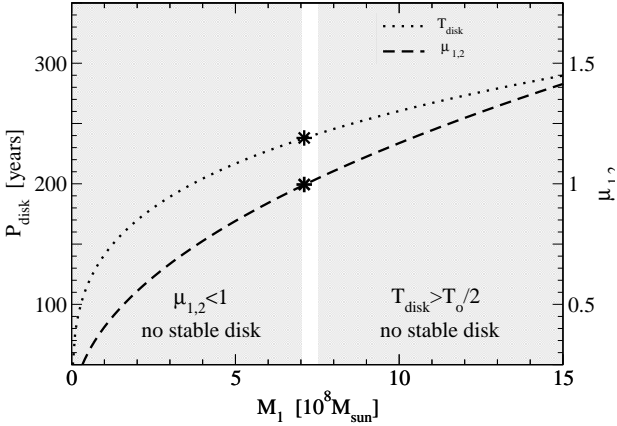}}
\caption{Range of acceptable solutions for the mass $M_1$ of the primary
black hole in \object{3C\,345}. The acceptable solutions exist within the range 
$7.1\times 10^8{\mathrm M}_{\sun} \le M_1 \le 7.5\times 10^8{\mathrm M}_{\sun}$.
Asterisks mark the solution for $\mu_\mathrm{ 1,2} = 1$, for which $M_1 =
M_2 = 7.1\times 10^8{\mathrm M}_{\sun}$. Shaded areas cover the values of $M_1$, $T_\mathrm{
o}$ and $\mu_{1,2}$ for which the accretion disk cannot remain stable
over the typical timescale of existence of a powerful radio source in AGN, $\tau_\mathrm{ agn}\le 10^8$\,yr.} 
\label{fg:sols-mass}
\end{figure}

\section{Discussion}
\label{sc:sect6}

The BBH model applied to the quasar \object{3C\,345} ties together, within a
single framework, the variability of the optical flux and the
evolution of the superluminal feature C7 resulted from a powerful flare
that occurred in 1991. The model reconstructs the physical conditions
in the plasma ejected into the jet during the flare, and the dynamic
properties of the binary system of supermassive black holes. 

The binary black hole scenario was previously applied to \object{3C\,345} by
Caproni \& Abraham (\cite{ca04}) who attributed variations of the ejection
angle (similar to those plotted in Fig.~\ref{fg:pa-change}) to
precession of the accretion disk. This approach yielded an
uncomfortably short precession period of $2.5\le T_\mathrm{ p}\le
3.8$\,yr, very high black hole masses ($M_1 = 4$--$5\times 10^9 {\mathrm
  M}_{\sun}$, $M_2 = 3$--$4\times 10^9 {\mathrm M}_{\sun}$), and
extremely small orbital separation $r_\mathrm{ o} = 0.017$--$0.024$\,pc,
which is $\approx 100\,R_\mathrm{ G}$ of the primary black hole. In such a
tight binary, the effective accretion radius of the secondary, $r_\mathrm{
  a} = 0.034$--$0.050$\,pc, is larger than the orbital separation, and
it is very difficult to maintain the accretion disk intact. Indeed, it
can be shown (Ivanov, Igumenshchev \& Novikov \cite{iin98}) that the mass loss
rate $\dot{m}_2$ due to a shock that is formed after each passage of
the secondary through the accretion disk may be very large, and for
the BBH parameters of Caproni \& Abraham (\cite{ca04}), it can reach
$\dot{m}_2 \approx \dot{M}_1 (\sqrt{10} \,\alpha_{\star}^{4/5})^{-1}$,
where $\dot{M}_1$ is the rate of accretion on the primary black hole
and $\alpha_{\star} = \alpha/10^{-2}$ is the disc viscosity.
For weak accretion rates or small disk viscosity, $\dot{m}_2$ can even
become larger than $\dot{M}_1$, which would most likely lead to a
subsequent destruction of the accretion disk.

Another argument against associating the ejection angle variability
with the disk precession is presented by the observed long-term
kinematic changes in the jet. The apparent accelerations plotted in
Fig.~\ref{fg:slopes} and the long-term trend present in the
ejection angle changes plotted in Fig.~\ref{fg:pa-change} indicate
the presence of a periodic process on timescales of $\sim
2000$\,yr. This observed periodicity, though very difficult to
establish, agrees well with the precession period recovered by fitting
the BBH model to the trajectory and flux evolution of C7. 

The trajectory of C7 (and indeed the trajectories of several other
superluminal features monitored closely in the jet of \object{3C\,345}; Lobanov
\& Zensus \cite{lz96}; \cite{L96}; \cite{LZ99}) is clearly
non-ballistic on timescales spanning almost 10 years in the observer's
frame (and more than 100 years in the comoving frame). This is a clear
observational argument against a very short precession period in
\object{3C\,345}. The model developed in this paper attributes non-ballistic
trajectories of the jet features to Alfv\'en perturbations of the
magnetic field, possibly modulated by the orbital motion in the
system.

\begin{figure}
\centerline{
\includegraphics[width=0.45\textwidth, bb = 15 58 690 610, clip=true]
{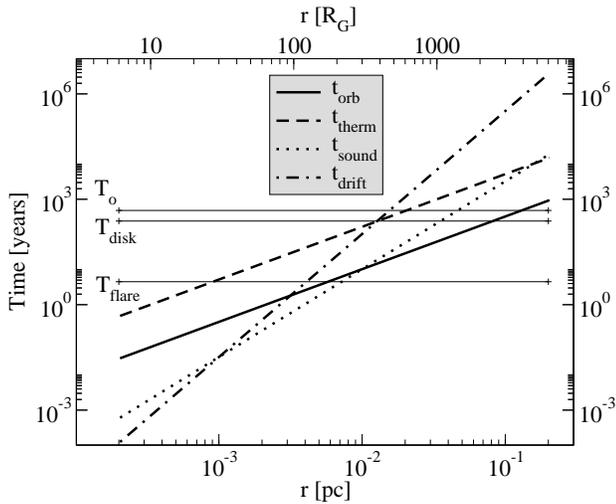}}
\caption{Characteristic timescales of disk activity in \object{3C\,345},
  compared to the quasi-period of the nuclear flares and to the
  orbital and characteristic disk rotation periods). The
   disk thermal instability operates at $t\sim 4$\,yr at distances of $\sim
  20\,R_\mathrm{ G}$. The mechanical (sound wave) instability reaches the
  same timescale at $\sim 200\,R_\mathrm{ G}$, where it becomes comparable
  with the orbital and drift (accretion rate variations) timescales of
  the disk. A combination of these effects may cause the flare
  activity observed in \object{3C\,345}.}
\label{fg:disk-scales}
\end{figure}

The sum of the black hole masses ($M_{1,2} = 1.4\times 10^9 {\mathrm
  M}_{\sun}$) estimated in our model agrees well with the mass
$M_\mathrm{ bh} = (4.0\pm 2.4)\pm 10^9{\mathrm M}_{\sun}$ obtained
from measurements of nuclear opacity and magnetic field strength in
\object{3C\,345} (Lobanov \cite{lob98a}). Gu, Cao \& Jiang (\cite{gcj01})
report a higher value of $M_\mathrm{ bh} = 7.6\times 10^9{\mathrm
  M}_{\sun}$ for \object{3C\,345}, from the properties of the broad line
region. This estimate, however, depend strongly on the monochromatic
luminosity ($M_\mathrm{ bh} \propto L_{5100{\mathrm \AA}}^2$) for
which the errors can be as large as 30\% even for the nearby objects
(Kaspi et al. \cite{kas+00}). This error could be even larger for weaker
objects (\object{3C\,345} is about 100 times weaker in the optical than an
average object from the sample of Kaspi et al. 2000), and
overestimating $L_{5100{\mathrm \AA}}^2$ by only 50\% would be bring
the $M_\mathrm{ bh}$ down to a value similar to the one derived in
this paper.
  
  Finally, let us consider the periodicity of the flares and component
  ejections in \object{3C\,345}. A $\sim 4$\,yr period of the flaring activity
  is inferred from the observed variability of the radio nucleus
  (\cite{LZ99}), and it is related with the ejections of superluminal
  features (see Fig.~\ref{fg:lcurve}). Since this period is much
  shorter than $T_\mathrm{ p} = 2570$\,yr, $T_\mathrm{ o} = 480$\,yr and
  $T_\mathrm{ disk} = 240$\,yr obtained in our model, there should be some
  other periodic process affecting the accretion disk on timescales of
  $\sim 4$\,yr. Figure~\ref{fg:disk-scales} shows four different
  characteristic disk timescales in \object{3C\,345}, for the black hole and
  disk parameters determined in Sect.~\ref{sc:sect4}. These
  timescales can be compared to the periodicity of the nuclear
  flares. It appears that several processes occurring in the disk at
  radial distances of 20--200$\,R_\mathrm{ G}$ may have similar
  characteristic timescales. Thermal instability at $\sim 20\,R_\mathrm{
    G}$ and mechanical (sound wave) instability at $\sim 200\,R_\mathrm{
    G}$ will develop at timescales of $\sim 4$\,yr. In addition to
  this, the orbital and drift (accretion rate variation) timescales
  approach $\sim 4$\,yr at $\sim 200\,R_\mathrm{ G}$. A combination of
  these effects can cause the observed flare activity.
  
  Another possible explanation for the observed flares is an
  intermediate or stellar mass black hole or even a neutron star
  orbiting the primary black hole at $r_{\star} \sim
  800-1200\,R_\mathrm{ G}$ at a large angle with respect to the
  accretion disk plane.  Close approaches to, or even passages
  through, the accretion disk should lead to formation of shock waves
  in the disk (Ivanov, Igumenshchev \& Novikov \cite{iin98}; \v{S}ubr
  \& Karas \cite{sk99}), which in turn would cause an increase in the
  accretion rate resulting in a nuclear flare and injection of a dense
  plasma cloud into the jet.
  
  The periodicity of the flares
  may also be caused by a turbulence in the jet. Formation of a
  turbulent layer close to the beam-jet boundary can enhance the pair
  creation rate in the beam, which would be observed as the appearance,
  and propagation, of a bright feature embedded in the flow.

\section{Conclusions}

The model developed in this paper explains the variability of the
optical flux and the kinematic and emission properties of the feature
C7 observed in the relativistic jet of \object{3C\,345} after a strong
nuclear flare. The dynamic properties of the system are described in
the framework of the orbital motion in a binary system of supermassive
black holes and the precession of an accretion disk around the primary
black hole. The emission and physical properties of relativistic
plasma produced during the flare are treated within the two-fluid
approach postulating a highly relativistic $e^{\pm}$ beam propagating
inside an $e^{-}p$ jet.  
 The $e^{-}p$ jet transports a major
  fraction of the kinetic power associated with the outflow from the
  nucleus of \object{3C\,345}, stabilizes
  the $e^{\pm}$ beam against plasma instabilities and shields it from
  interaction with the external medium.

Determination of the properties of the jet and the binary system in
\object{3C\,345} is a difficult process involving two major steps. First, the
precession characteristics of the accretion disk are determined by
fitting the kinematic and flux density evolution of C7. This yields a
precession period $T_\mathrm{ p} = 2570$\,yr, which is in a good
agreement with the periodicities inferred from the changes of
kinematic properties of individual jet components and variations of
the ejecting position angle in \object{3C\,345}.

In the second step, the
orbital and physical properties of the binary system are recovered
by fitting the observed optical variability. This yields an orbital
period of $T_\mathrm{ p} = 480$. The consistency between
the two fits is ensured by requiring the fit to the optical light
curve to reproduce also the kinematic and flux density evolution of
C7. The consistency argument constrains the properties of radio and
optically emitting plasma, with $\Gamma_\mathrm{ min, opt} = 12$ and
$\Gamma_\mathrm{ min,rad} = 4$. 

After ensuring consistency between the two fits, they are combined in
order to constrain the Alfv\'en speed $V_\mathrm{ A} = 0.078\,c$, and
make it popsible to recover the physical properties of the system.  In
the last step, the conditions of disk stability are employed to
determine the individual masses of the binary companions.  This
procedure yields $M_1 = M_2 = 7.1\times 10^8{\mathrm M}_{\sun}$.
Finally, the model offers an explanation for the observed 4-year quasi
periodicity of nuclear flares in \object{3C\,345}. This
quasi-periodicity arises naturally from the characteristic timescales
of the accretion disk surrounding the primary black hole.  The overall
success that the model has in explaining a variety of the
observational properties in \object{3C\,345} argues in favor of this
explanation.



\begin{thebibliography}{99}


\bibitem[1998]{ac98} Abraham, Z. \& Carrara,
  E.~A. 1998, ApJ, 496, 172 

\bibitem[1999]{ar99} Abraham, Z. \& Romero,
  G.~E. 1999, A\&A, 344, 61

\bibitem[1993]{as93} Achatz, U. \& Schlickeiser, R., 
  1993, A\&A, 274, 165
  
\bibitem[2003]{aller03} Aller, M.~F., Aller,
  H.~D. \& Hughes, P.~A.\ 2003, ApJ, 586, 33

\bibitem[1981]{ang+81} Angione, R.~J., Moore, E.~P.,
  Roosen, R.~G. \& Sievers, J. 1981, AJ, 86, 653

\bibitem[1999]{arw99} Attridge, J.~M., Roberts, D.~H., 
  \& Wardle, J.~F.~C. 1999, ApJ, 518, 87
  
\bibitem[1992]{baa+92} B{\aa}{\aa}th, L.~B. 
  Rogers, A.~E.~E., Inoue, M., et al. 1992, A\&A, 257, 31

\bibitem[1984]{bb84} Babadzhanyants, M.~K., 
\& Belokon', E.~T. 1984, Astrophysics, 20, 461

\bibitem[1995]{bbg95} Babadzhanyants, M.~K., 
Belokon', E.~T. \& Gamm, N.~G. 1995, Astronomy Reports, 39, 393

\bibitem[1999] {bb99} Belokon, E.~T. 
\& Babadzhanyants, M.~K. 1999,  Astronomy Letters,  25,  781

\bibitem[1980]{bbr80} Begelman, M.~C., 
  Blandford, R.~D. \& Rees, M.~J. 1980, Nature, 287, 307
  
\bibitem[1986]{bmc86} Biretta, J.~A., Moore, R.~L. \&
  Cohen, M.~H. 1986, ApJ, 308, 93
  
\bibitem[1999]{bsm99} Biretta, J.~A., Sparks, W.~B.
  \& Macchetto, F. 1999, ApJ, 520, 621

\bibitem[1978]{br78} Blandford, R.~D. \& Rees, M.~J.
  1978, in Pittsburgh Conference on BL Lac Objects, ed. A.~M.~Wolfe
  (Pittsburgh: University of Pittsburgh), 328
  
\bibitem[2001] {bri+01} Britzen, S., Roland, J., Laskar, J. et al.
  2001, A\&A, 374, 784

\bibitem[2004]{ca04} Caproni, A. \& Abraham,
  Z. 2004, ApJ, 602, 625

\bibitem[1992]{ck92} Camenzind, M. \&
  Krockenberger, M. 1992, A\&A, 255, 59

\bibitem[1985]{coh85} Cohen, M.~H. 1985, in Extragalactic Energetic
Sources, ed. V.~K.~Kapahi (Bangalore: Indian Academy of Sciences), 1

\bibitem[1997] {df97} Despringre, V. \&
  Fraix-Burnet, D. 1997, A\&A, 320, 26

\bibitem[1993]{fer+93}  Feretti, L., Comoretto, G., 
  Giovannini, G., et al. 1993, ApJ, 408, 446

\bibitem[2001]{gcj01} Gu, M., Cao, X. \& Jiang, D.~R. \
  2001, MNRAS, 1111

\bibitem[1996] {hs96} Hanasz, M. \& Sol, H.  1996, A\&A,
  315, 355

\bibitem[1993]{hb93} Hewitt, A. \& Burbidge, G. 1993, 
ApJS, 87, 451

\bibitem[1998]{iin98} Ivanov, P.~B.,
  Igumenshchev, I.~V. \& Novikov, I.~D. 1998, ApJ, 507, 131

\bibitem[1999] {ipp99} Ivanov, P.~B., 
  Papaloizou, J.~C.~B. \& Polnarev, A.~G. 1999, MNRAS,  307,  79

\bibitem[2001]{jor+01} Jorstad S.~G., Marscher A.~P., 
  Mattox J.~R. et al. 2001, ApJ,  556, 738 

\bibitem[2001a]{jor+01a} Jorstad S.~G., Marscher A.~P., 
  Mattox J.~R. et al. 2001, ApJS,  134, 181 

\bibitem[1992]{kr92} Kaastra, J.~S. \& Roos, N. 1992,
  A\&A, 254, 96
  
\bibitem[2000]{kas+00} Kaspi, S., Smith, P.~S., Netzer et al.\ 2000,
  ApJ, 533, 631

\bibitem[1988]{kid88} Kidger, M.~R. 1988, PASP, 100, 1248

\bibitem[1989]{kid89} Kidger, M.~R. 1989, A\&A, 226, 9
  
\bibitem[990 ]{kd90} Kidger, M.~R.~\& de Diego,
  J.~A. 1990, A\&A, 227, L25

\bibitem[1968]{kin+68} Kinman, T.~D., Lamla, 
E., Ciurla, T., Harlan, E., \& Wirtanen, C.~A. 1968, ApJ, 152, 357 

\bibitem[2003]{kla03} Klare, J. 2003, PhD Thesis, University
  of Bonn, Germany

\bibitem[1989]{kwr89} Kolgaard, R.~I., Wardle,
  J.~F. \& Roberts, D.~H. 1989, ApJ, 97, 1550

\bibitem[1981]{kon81} K\"onigl, A. 1981, ApJ, 243, 700

\bibitem[2002]{lb02} Laing, R. \& Bridle,
  A.~H. 2002, MNRAS, 336, 1161

\bibitem[2004]{lb04} Laing, R. \& Bridle,
  A.~H. 2004, MNRAS, 348, 1459

\bibitem[1990]{las90} Laskar, J. 1990, in ``M\'{e}thodes Modernes de 
  la M\'{e}canique C\'{e}leste'', eds. D. Benest, C. Froeschl\'{e}, 89

\bibitem[1995]{lr95} Laskar J. \& Robutel P. 1995, 
  Celestial Mechanics, 62, 193

\bibitem[1996]{lv96} Lehto, H.~J., \& Valtonen, M.~J. 1996, 
  ApJ 460, 270

\bibitem[1995]{lzd95} Lepp\"anen, K.~J., Zensus,
  J.~A., \& Diamond, P.~J. 1995, AJ, 110, 2479

\bibitem[L96]{L96} Lobanov, A.~P. 1996, PhD Thesis, NMIMT, Socorro,
  USA (L96)

\bibitem[1998a]{lob98a} Lobanov, A.~P. 1998a,  A\&A, 330, 79

\bibitem[1998b]{lob98b} Lobanov, A.~P. 1998b, A\&AS, 133, 261
  
\bibitem[1997] {lcz97} Lobanov, A.~P.,
  Carrara, E. \& Zensus, J.~A. 1997, Vistas in Astronomy, 41, 253
  
\bibitem[1996]{lz96} Lobanov, A.~P. \& Zensus, J.~A.
  1996, in ASP Conf. Ser. v.\,100, Energy Transport in Radio Galaxies and
  Quasars, ed.  P.~E.~Hardee, J.~A.~Zensus \& A.~H.~Bridle (San
  Francisco: ASP), 124

\bibitem[LZ99]{LZ99} Lobanov, A.~P. \& Zensus, J.~A. 1999,
  ApJ, 521, 509 (LZ99)

\bibitem[1972]{lu72} L\"u, P.~K. 1972, AJ, 77, 829 

\bibitem[1997] {mak97} Makino, J. 1997, ApJ,  478,  58 

\bibitem[1996] {me96} Makino, J. \&  Ebisuzaki, T. 1996,
  ApJ,  465,  527 

\bibitem[1995]{mhp95} Marcowith, A., Henri, G. 
  \& Pelletier, G. 1995, MNRAS, 277, 681

\bibitem[1998]{mhr98} Marcowith, A., Henri, G., 
  \& Renaud N. 1998, A\&A, 331, L57
  
\bibitem[1975]{mcg+75} McGimsey, B.~Q., Smith, A.~G.,
  Scott, R.~L., et al. 1975, AJ, 80, 895


\bibitem[1988]{psa88} Pelletier, G., Sol,
  H. \& Asseo, E., 1988, Phys. Rev. A 38, 2552

\bibitem[1989]{pr89} Pelletier, G. \& Roland, J.
  1989, A\&A, 224, 24
  
\bibitem[1990]{pr90} Pelletier, G., Roland, J.,
  1990. In: ``Parsec-Scale Jets'', eds. J.~A. Zensus \& T.~J. Pearson,
  (Cambridge University Press: Cambridge), 323

\bibitem[1992] {ps92} Pelletier, G. \& Sol, H. 1992,  
  MNRAS,  254,  635
  
\bibitem[1979]{pol+79} Pollock, J.~T., Pica, A.~J., Smith, A.~G., et
  al.\ 1979, AJ, 84, 1658

\bibitem[1994]{pr94} Polnarev, A.~G. \& Rees, M.~J. 1994,
  A\&A, 283, 301

\bibitem[2000] {rm00} Rieger, F.~M. \& Mannheim, K. 2000,   
  A\&A,  359,  948

\bibitem[1995]{rh95} Roland J. \& Hermsen W. 1995, 
  A\&A, 297, L9
  
\bibitem[1996]{rh96} Roland, J. \& Hetem, A., 1996,
  in: Cygnus A: Study of a radio galaxy, eds. C. L.
  Carilli \& D. E. Harris, (Cambridge University Press,
  Cambridge), 126

\bibitem[1988]{rpm88} Roland, J., Peletier, G. \& Muxlow, T. 1988,
  A\&A, 207, 16

\bibitem[1994]{rtr94} Roland J., Teyssier R. \& 
  Roos N. 1994, A\&A, 290, 357

\bibitem[1993] {rkh93} Roos, N., Kaastra, J.~S. \& 
  Hummel, C.~A. 1993,  ApJ,  409,  130

\bibitem[R00]{R00} Ros, E., Zensus, J.~A. \& 
  Lobanov, A.~P. 2000, A\&A, 354, 55 (R00)

\bibitem[1988]{shv88}Sillanp\"a\"a, A., 
  Haarala, S. \& Valtonen, M.~J. 1988, ApJ, 325, 628
  
\bibitem[1993]{sch+93} Schramm, K.-J., Borgeest, U.,
  Camenzind, M., et al. 1993, A\&A, 278, 391

\bibitem[1997]{sds97} Skibo, J.G., Dermer, C.D. 
  \& Schlickeiser, R. 1997, ApJ, 483, 56

\bibitem[1970]{sw70} Smyth, M.~J. \& 
Wolstencroft, R.~D.  1970, A\&ASS, 8, 471 

\bibitem[1989]{spa89} Sol H., Pelletier G. \& 
  Asseo E. 1989, MNRAS, 237, 411

\bibitem[1995]{ste+95} Steffen, W., Zensus, J.~A., 
Krichbaum, T.~P., Witzel, A. \& Qian, S.~J. 1995, A\&A, 302, 335

\bibitem[1999]{sk99} \v{S}ubr, L. \& Karas, V. \
  1999, A\&A 352, 452

\bibitem[1998]{ter+98} Ter\"asranta, H.,
  Tornikoski, M., Mujunen, A., et al. 1998, A\&AS, 132, 305
  
\bibitem[1998]{tin+98} Tingay, S.~J., Jauncey, D.~L.,
  Reynolds, J.~E., et al. 1998, AJ, 115, 960

\bibitem[2000]{val+00} Valtaoja, E., Ter\"asranta, H.,
  Tornikoski, M. et al. 2000, ApJ, 531, 744

\bibitem[1996]{vcs96} Vicente, L., Charlot, P. \& 
  Sol, H. 1996, A\&A 312, 727

\bibitem[1991]{vio+91} Vio, R., Cristiani, S., Lessi, O. \& 
Salvadori, L. 1991, ApJ, 380, 351

\bibitem[1983]{unw+83} Unwin, S.~C., Cohen, M.~H.,
  Pearson, T.~J., et al., 1983, ApJ, 271, 536
  
\bibitem[1994]{unw+94} Unwin, S.~C., Wehrle, A.~E., Urry,
  C.~M., et al., 1994, ApJ, 432, 103
  
\bibitem[1997] {unw+97} Unwin, S.~C., Wehrle, A.~E.,
  Lobanov, A~.P., et al., 1997, ApJ, 480, 596

\bibitem[1991]{wal+91} Waltman, E.~B., Fiedler, R.~L.,
  Johnston, K.~J., et al. 1991, ApJS, 77, 379
  
\bibitem[1988]{web+88} Webb, J.~R., Smith, A.~G., Leacock, R.~J., et
  al.\ 1988, AJ, 95, 374

\bibitem[1997]{zen97} Zensus, J.~A. 1997, ARA\&A, 35, 607

\bibitem[1995]{zkl95} Zensus, J.~A., Krichbaum, T.~P. \& Lobanov,
  A.~P. 1995, Proc. Natl. Acad. Sci. USA, 92, 11348

\bibitem[1996]{zkl96} Zensus, J.~A., Krichbaum, T.~P. \& Lobanov,
  A.~P.\ 1996, Rev. Mod. Astron., 9, 241

\bibitem[ZCU95]{ZCU95} Zensus, J.~A., Cohen,
  M.~H.  \& Unwin, S.~C. 1995, ApJ, 443, 35 (ZCU95)

\bibitem[2002]{zen02} Zensus, J.~A., Ros, E.,
  Kellermann, K.~I., Cohen, M.~H., Vermeulen, R.~C. \& Kadler,
  M. 2002, AJ, 124, 662


\end{thebibliography}
\end{document}